\documentclass[twocolumn,trackchanges]{aastex631}

\newcommand\soutm{\bgroup\markoverwith{\textcolor{black}{\rule[0.5ex]{2pt}{0.8pt}}}\ULon}

\shorttitle{GCs in the Virgo cluster}
\shortauthors{Park et al.}
\graphicspath{{./}{figures/}}

\usepackage{CJK}
\usepackage{ulem}

\begin{document}
\begin{CJK*}{UTF8}{}
\CJKfamily{mj}

\title{Properties of Globular Clusters in Galaxy Clusters: Sensitivity from the Formation and Evolution of Globular Clusters}

\correspondingauthor{Jihye Shin}
\email{jhshin@kasi.re.kr}

\author[0000-0003-1889-325X]{So-Myoung Park}
\affiliation{Korea Astronomy and Space Science Institute \\
776 Daedeok-daero, Yuseong-gu \\
Daejeon 34055, South Korea}

\author[0000-0001-5135-1693]{Jihye Shin}
\affiliation{Korea Astronomy and Space Science Institute \\
776 Daedeok-daero, Yuseong-gu \\
Daejeon 34055, South Korea}

\author[0000-0001-5303-6830]{Rory Smith}
\affiliation{Universidad T\'{e}cnica Federico Santa Mar\'{i}a \\
3939 Vicu\~{n}a Mackenna, San Joaqu\'{i}n \\
Santiago 8940897, Chile}

\author[0000-0001-9544-7021]{Kyungwon Chun}
\affiliation{Korea Astronomy and Space Science Institute \\
776 Daedeok-daero, Yuseong-gu \\
Daejeon 34055, South Korea}



\begin{abstract}

We investigate the properties of globular clusters in a galaxy cluster, using the particle tagging method with a semi-analytical approach in a cosmological context.
We assume globular clusters form from dark matter halo mergers and their metallicity is assigned based on the stellar mass of the host dark matter halos and the formation redshift of GCs.
Dynamical evolution and disruption of globular clusters are considered using semi-analytical approaches, controlled by several free parameters.
In this paper, we investigate how our results are changed by the choice of free parameters.
We compare our fiducial results with representative observations, including the mass ratio between the globular cluster system and its host galaxy, the globular cluster occupancy, the number fraction of blue globular clusters, and the metallicity gradient with the globular cluster mass.
Because we can know the positions of globular clusters with time, comparison with additional observations is possible, e.g., the median radii of the globular cluster system in individual galaxies, the mean projected density profiles of intracluster globular clusters, and metallicity and age gradients of globular clusters with a clustercentric radius.
We also find that the specific mass of the globular cluster system in each galaxy is different with a clustercentric radius.

\end{abstract}



\section{Introduction}
\label{sec1}

Globular clusters (GCs) are found in every type of galaxy, from dwarfs to cD galaxies.
Their typical mass range is $10^{4}$-$10^{6}$~M$_{\odot}$ and their present mass function (MF) shows a log-normal shape \citep{Fall+2001,Waters+2006,Jordan+2007b,Lomeli-Nunez+2022}.
Because their ages are typically nearly a Hubble time ago ($\sim$10-12 Gyrs), they are fossils that represent the extreme star formation at high redshift \citep[e.g.][]{Glatt+2008,Chies-Santos+2011,Powalka+2017,Usher+2019}.
GCs in galaxies and galaxy clusters provide valuable information about the merging history of galaxies and galaxy clusters and the environment of GC formation \citep{Mackey+2010,Olchanski+2018,Kruijssen+2019b,Massari+2019,Dolfi+2021}.
Therefore, GCs are useful to understand the formation and evolution of their host galaxies \citep{Brodie+2006,Forbes+2018a}.

GC observations have been conducted in local groups and galaxy clusters \citep{Cote+2004,Jordan+2007a,Sarajedini+2007,Carter+2008,Ferrarese+2012,Harris+2013,Brodie+2014,Piotto+2015,Fahrion+2020} and they have revealed important relations between GCs and their host galaxies.
Basically, most galaxies of $M_{\rm stellar} \geq 10^{9}$~M$_{\odot}$ have GCs \citep[GC occupancy:][]{Peng+2008,Georgiev+2010,Sanchez-Janssen+2019,Carlsten+2021}, where $M_{\rm stellar}$ is the galaxy stellar mass.
Although there is a non-linear relation between the total and stellar mass of galaxies \citep{Behroozi+2013a,Hudson+2015}, galaxies including GCs show a constant mass fraction at $z=0$ of $M_{\rm GCs}/M_{\rm halo} \sim 5 \times 10^{-5}$ \citep[the $M_{\rm GCs}$-$M_{\rm halo}$ relation:][]{Peng+2008,Harris+2013,Hudson+2014,Harris+2015}, where $M_{\rm GCs}$ is a GC system mass in each galaxy, and $M_{\rm halo}$ is the total galaxy mass.
Previous papers have suggested that the $M_{\rm GCs}$-$M_{\rm halo}$ relation is the result of hierarchical galaxy mergers \citep[e.g.,][]{El-Badry+2019,Choksi+2019b,Bastian+2020}.

GCs in galaxies exhibit color bimodality, which is commonly observed in almost all massive early-type galaxies in Virgo itself but is not universally observed \citep{Larsen+2001,Forbes2005,Harris+2006,Peng+2006,Waters+2009,Brodie+2012,Tonini+2013,Harris+2016,Bastian+2018}.
Blue GCs are metal-poor, while red GCs are metal-rich.
Color bimodality of GCs might imply that there were two star formation epochs or mechanisms in the history of galaxies \citep{Brodie+2006,Usher+2012}.

Based on color bimodality of GCs, there are three different formation scenarios of blue and red GCs \citep[see][for a review]{Brodie+2006}: the major merger model, the in-situ scenario, and the dissipationless accretion scenario.
The major merger model is for when blue GCs form in protogalactic fragments, while red GCs form from the gas-rich major merger of galaxies \citep{Ashman+1992}.
The in-situ scenario is when blue GCs form simultaneously as galaxies form but the formation is restrained by the pressure from supernova explosion that causes gas to be expelled from the star-forming galaxy.
Then, red GCs form from the expelled gas, which falls back into the galaxy due to gravity \citep{Forbes+1997,Harris+1999}.
Finally, in the dissipationless accretion scenario, blue GCs form by the dissipationless accretion of neighbouring dwarf galaxies and red GCs form in-situ in massive seed galaxies \citep{Cote+1998}.

On the other hand, there is an alternative GC formation scenario, based on Lambda cold dark matter ($\Lambda$CDM) cosmology: the hierarchical merging scenario \citep{Muratov+2010}.
In this scenario, the hierarchical build-up of galaxies by mergers makes GCs and one does not need to consider the formation of blue and red GCs independently.
Instead, the metallicity of GCs is assigned by the $M_{\rm stellar}$ of their host galaxies and formation redshift of GCs.

To trace the formation and evolution of GCs, various numerical simulations have been performed.
Due to the wide dynamical range between GCs and their host galaxies, high-resolution cosmological simulations are required.
Because of the huge advancements in computational hardware, recently high-resolution cosmological hydrodynamic simulations have been performed so they have started to directly resolve the GC formation with somewhat limited internal resolution \citep[][and references therein]{Boley+2009,Kimm+2016,Li+2017,Kim+2018,Lahen+2020,Ma+2020}.
However, they still suffer from the limitations of spatial/mass resolution, simulation volume, and huge calculation times.

As an extension to the approach of the high-resolution cosmological hydrodynamic simulations, the {\sc e-mosaics} project performs a self-consistent hydrodynamical zoom-in simulation with subgrid modelling \citep{Pfeffer+2018,Kruijssen+2019a,Reina-Campos+2022a,Reina-Campos+2022b}.
In this hydrodynamical approach, we can infer the formation site and condition of the GC formation.
The E-MOSAICS project applies the GC mass loss due to stellar evolution, tidal shock, two-body relaxation, and dynamical friction, but only dynamical friction is calculated in post-processing.
It increases the calculation time of modelling so there is a limitation to investigate the properties of GCs by exploring the impact of parameter variations.

Traditionally, DM-only simulations with a semi-analytical approach for the GC formation are implemented.
DM-only simulations require a short computational time, but generally they do not make clear predictions on the GC spatial distributions \citep{Prieto+2008,Li+2014,Choksi+2018,El-Badry+2019}.

The particle tagging method (PTM) that tags DM or stellar particles as GCs also has been applied in cosmological DM-only or hydrodynamical simulations \citep{Corbett_Moran+2014,Mistani+2016,Ramos-Almendares+2018,Ramos-Almendares+2020,Chen+2022}.
The advantages of the PTM are that we can trace the position and velocity information of GCs and the computational time is much shorter than high-resolution cosmological simulations due to the post-processing and without running full simulations.
Previous studies using this method, however, do not consider various processes of dynamical evolution of GCs.
For example, \citet{Ramos-Almendares+2018,Ramos-Almendares+2020} neglect the mass-loss process of GCs to concentrate on the tidal stripping that causes GCs to escape their host galaxies and become intraclutser GCs \citep[ICGCs,][]{Lee+2010}.
\citet{Chen+2022} apply the stellar and tidal mass-loss of GCs but neglect the mass-loss by two-body relaxation, dynamical fraction, and tidal shocks.

In this paper, we investigate the properties of GCs in galaxy clusters, combining advantages of both PTM and a semi-analytical approach: the position information from PTM and the speed from a semi-analytical approach.
Based on the DM-only cosmological simulation, we tag DM particles as GCs and apply the dynamical evolution of each GC.
For the GC formation, we adopt the hierarchical merging scenario, which assumes GCs form from DM halo mergers and the metallicity of GCs is a function of $M_{\rm stellar}$ of their host DM halos and GC formation redshift.
The GC evolution are described by the semi-analytical approach so it can be easily adjusted by varying the parameter sets that control the analytical recipes.
We investigate how parameter variations can affect the properties of GCs with the advantage of the speed of our method and then we compare our results with various observations, including position information.
In this study, first, we introduce our method optimized for GC formation and evolution.
Second, we take advantage of the speed at which we can explore parameter space with our method.
This enables us to investigate the sensitivity of our results to possible parameters, allowing us to better understand which physical processes are important.
Finally, we compare our fiducial results with various observations.
Note that our main purpose is not to tune our fiducial parameter set to reproduce the observations exactly.

This paper is constructed as follows.
In Section \ref{sec2}, we introduce our PTM and show how GCs form and evolve in a cosmological context with the semi-  analytical approach.
Section \ref{sec3} shows the comparison our results with observations and how the choice of parameter set can affect the results.
In Section \ref{sec4}, we compare our results with additional observations.
Sections \ref{sec5} and \ref{sec6} discuss and summarize our results.

\section{Method for globular cluster formation and evolution}
\label{sec2}

In this section, we will introduce our PTM with the semi-analytical approach to describe the formation and evolution of GCs.
For the GC formation, we adopt the hierarchical merging scenario \citep[e.g.][and references therein]{Choksi+2018,El-Badry+2019,Choksi+2019a,Bastian+2020}.
For the GC mass-loss, we apply the stellar evolution and two-body relaxation.
For the GC disruption, we apply tides from host galaxies and the orbital decay by dynamical friction.

\subsection{Halo mass assembly history}

We build the cluster mass assembly history using 3 sets of high-resolution DM-only cosmological simulations from $z=200$ to $z=0$ \citep{Taylor+2019} with the $N$-body code, {\sc gadget2} \citep{Springel2005}.
The original simulations were carried out with cosmological parameters $\Omega_{m}=0.3$, $\Omega_{\Lambda}=0.7$, $\Omega_{b}=0.04$, and $h=0.68$ with a box of (140~Mpc/$h$)$^{3}$.
The mass of each particle is $1.7\times10^{9}$~M$_{\odot}/h$ with a softening length ($\epsilon$) of 5.469~kpc/$h$.
The power spectrum is calculated by the {\sc camb} package\footnote{\url{http://camb.info}}\citep{Lewis+2000}.

From these original simulations, Virgo cluster analogue DM halos\footnote{The Virgo cluster is one of the the nearest galaxy clusters so various kinds of observations are implemented: the ACS Virgo cluster survey \citep{Cote+2004}, the next generation Virgo cluster survey (NGVS; \citealt{Ferrarese+2012}), and the Extended Virgo Cluster Catalog (EVCC; \citealt{Kim+2014}), and so on. Therefore, we can use various observational properties of GCs in the Virgo cluster.} are identified at $z=0$.
We choose three cluster DM halos (Target 1-3), which have small Lagrangian volumes, among various Virgo cluster analogue DM halos in the original simulations to reduce the calculation time.
Next, we resimulate Targets 1-3 with a zoom-in technique, which involves rerunning the interesting regions of the low-resolution simulations with higher resolution \citep{Porter1985,Katz+1993,Navarro+1994}.
Multiscale initial conditions with positions and velocities were generated by the {\sc music} package\footnote{\url{https://www-n.oca.eu/ohahn/MUSIC/}} \citep{Hahn+2011}.
The high-resolution particle mass is 3.32~$\times~10^{6}$~M$_{\odot}/h$ with $\epsilon=0.683$~kpc/$h$.
Targets 1-3 have a $M_{\rm halo}$ of 9.25$\times10^{13}$, 1.14$\times10^{14}$, and 1.26$\times10^{14}$~M$_{\odot}/h$, respectively.
The virial ratio, $2T/|U|$, of Target 1-3 is 1.56, 1.19, and 1.31, respectively, where $T$ is a kinetic energy and $U$ is a potential energy so the Target 1 is the most unrelaxed cluster among them.

We identify halo and subhalo structures with the modified six-dimensional phase-space DM halo finder, {\sc rockstar}\footnote{\url{https://bitbucket.org/pbehroozi/rockstar-galaxies}} \citep{Behroozi+2013b} and make merger trees using Consistent Trees \citep{Behroozi+2013c}.
To define the DM halo in the {\sc rockstar}, we set the minimum number to 20 particles so the minimum $M_{\rm halo}$ is $6.64\times10^{7}$~M$_{\odot}/h$.
Throughout the paper, however, we use DM halos more massive than $M_{\rm halo}$ of $10^{9}$~M$_{\odot}/h$, which consists of more than $\sim$300 DM particles.
The equivalent $M_{\rm stellar}$ of $M_{\rm halo}\simeq10^{9}$~M$_{\odot}/h$ would be 5.97$\times10^{5}$~M$_{\odot}/h$.
Here, We convert from $M_{\rm halo}$ to $M_{\rm stellar}$, using the $M_{\rm stellar}$ and $M_{\rm halo}$ relation from weak lensing data \citep{Hudson+2015}.
This match is what was done in the observation so we can compare our results with the observations identically.
The slope of stellar MF of the observation is -0.56 \citep{Lan+2016}, and Targets 1-3 have slopes of stellar MF with -0.54, -0.54, and -0.57, respectively.

\subsection{Particle tagging method for globular clusters}
\label{sec2.2}

The hierarchical merging scenario is adopted to make GCs in our model.
We assume GCs form by DM halo mergers that are larger than a minimum mass ratio of $\gamma_{\rm MR}$, which is a free parameter in our model.
When GCs form, the initial $M_{\rm GCs}$ in each DM halo has a linear relation with $M_{\rm halo}$ \citep{Peng+2008,Harris+2013,Hudson+2014,Harris+2015}:
\begin{equation}
    M_{\rm GCs} = \eta M_{\rm halo},
\end{equation}
where $\eta$ is a constant formation efficiency and a free parameter in our model.

We assume that one DM particle represents one GC, although the mass of the GC is not equal to the mass of the DM particle.
The initial $M_{\rm GCs}$ is distributed into individual GCs with a power-law initial GCMF \citep{Elmegreen2018}:
\begin{equation}
    \label{eq2}
    \frac{{\rm d}N}{{\rm d}M}=M_{0}M^{-2},
\end{equation}
where $M_{0}$ is a normalisation constant.
We determine the minimum mass of the GC ($M_{\rm min}$) is $10^{5}$~M$_{\odot}$ because we assume that most GCs below $10^{5}$~M$_{\odot}$ are destroyed due to two-body relaxation.
If we normalize the probability of the MF, Equation (\ref{eq2}) is
\begin{equation}
    1 = \int_{M_{\rm min}}^{\infty}M_{0}M^{-2}{\rm d}M,
\end{equation}
which gives $M_{0} = M_{\rm min}$.
Then, using the transformation method, the initial mass of each GC, $M_{i}$, is selected by a random number, $0<r<1$:
\begin{equation}
    \label{eq4}
    M_{i} = \frac{M_{\rm min}}{r}.
\end{equation}
We repeat Equation (\ref{eq4}) until the sum of the initial mass of each GC in individual DM halos reaches the required $M_{\rm GCs}$ value at GC formation epoch and, in this way, the number of GCs is also determined.
The same number of DM particles from each DM halo is then selected to trace out the location of the GCs.
Here, we pick particles according to in order of their bound energies, which assumes GCs form in the deepest location of the potential potential of DM halos.
Finally, a different GC initial mass is assigned to each of the tagged DM particles, allowing us to trace their position and velocity down to $z=0$.

We assume that the metallicity of GCs is assigned by $M_{\rm stellar}$ of their host DM halos and redshift at GC formation epoch \citep{Li+2014,Choksi+2018,Choksi+2019a}.
We adopt the stellar mass-metallicity model \citep{Ma+2016,Choksi+2018}:
\begin{equation}
    \label{eq5}
    [{\rm Fe/H}] = \log\Bigg\{
    \left(\frac{M_{\rm stellar}}{10^{10.5}~{\rm M}_{\odot}}\right)^{\alpha_{m}}
    (1+z)^{-\alpha_{z}}
    \Bigg\},
\end{equation}
where $\alpha_{m}=0.35$ and $\alpha_{z}=0.9$ \citep[e.g.,][]{Choksi+2018,Chen+2022}.
Here, we define metal-poor (blue) GCs as [Fe/H]$<$-1.0 and metal-rich (red) GCs as [Fe/H]$>$-1.0.

\subsection{Mass evolution and disruption of globular clusters}

After GCs form, they undergo mass-loss and disruption by several internal and external processes: stellar evolution, two-body relaxation, tides from host galaxies, the orbital decay by dynamical friction, and tidal shocks \citep[e.g.,][and references therein]{Spitzer1987,Gieles+2011,Shin+2013,Webb+2014,Madrid+2017}.
We include recipes for these processes in our PTM with the seim-analytical approach.
Note that we do not consider the GC disruption by tidal shocks because our simulations are DM-only simulations and we do not model the gas and stellar distributions in each galaxy.

\subsubsection{Mass loss of GCs by stellar evolution and two-body relaxation}

The GC mass evolution is approximated to the first-order differential equation \citep{Fall+2001}:
\begin{equation}
    \frac{{\rm d}M}{{\rm d}t} = -(\nu_{\rm se}+\nu_{\rm rlx})M_{i},
\end{equation}
where $\nu_{\rm se}$ and $\nu_{\rm rlx}$ are time-dependent fractional mass-loss rates by the stellar evolution and two-body relaxation, respectively.

For $\nu_{\rm se}$, we use the stellar evolution model \citep{Hurley+2000}, assuming the initial mass function (IMF) of each GC is distributed by Kroupa IMF \citep{Kroupa2002}.
We tabulate the changes in the mass of stars with different masses as a function of time, $\nu_{\rm se}(t)$, so we can calculate the amount of mass-loss of each GC after they form.
We assume a constant metallicity (0.1~Z$_{\odot}$ ), which is a typical metallicity of observed GCs\footnote{For simplicity, we do not use assigned the metallicity by Equation (\ref{eq5}).}.
Figure \ref{fig1} shows the evolution of GCMF in the brightest cluster galaxy (BCG).
The black thin histogram shows the MF without a mass-loss and disruption of GCs (original MF), and the green histogram shows the evolved MF, where we only apply the mass-loss by the stellar evolution.
The stellar evolution does not change the shape of the MF but it moves the original MF to the low-mass part.

For $\nu_{\rm rlx}$, we adopt a formula in \citet{Spitzer1987}:
\begin{equation}
    \label{eq7}
    \nu_{\rm rlx} = \frac{\xi_{e}}{t_{\rm rh}},
\end{equation}
where $\xi_{e}$ of 0.033 is the normalization factor and $t_{\rm rh}$ is the half-mass relaxation time scale:
\begin{equation}
    \label{eq8}
    t_{\rm rh} = \frac{M_{i}^{1/2}r_{\rm h0}^{3/2}}
                        {7.25\bar{m}G^{1/2}\ln{\Lambda}},
\end{equation}
whee $\bar{m}$ of 0.87~M$_{\odot}$ is the mean stellar mass of a Kroupa IMF, $\ln{\Lambda}$ of 12 is the Coulomb logarithm, a typical value for GCs, and $r_{\rm h0}$ is the initial half-mass radius of GCs.
We treat $r_{\rm h0}$ as a free parameter in our model.

The blue histogram in Figure \ref{fig1} shows the evolved MF after applying the mass-loss by the stellar evolution and two-body relaxation.
The shape of the original MF is changed to a log-normal shape by two-body relaxation.

\subsubsection{Disruption of GCs by tides and dynamical friction}

To consider tides from host DM halos where GCs belong, we divided GCs into those in the `isolated regime' and those in the `tidal regime', using $\Re=r_{\rm h}(t)/r_{\rm J}(t)$, where $r_{\rm h}(t)$ is a half-mass radius and $r_{\rm J}(t)$ is the Jacobi radius\footnote{For simplicity, we assume the tidal radius is the Jacobi radius throughout the paper.} with time \citep{Gieles+2008}.
The Jacobi radius is the distance from the cluster center to the Lagrange points $L_{1}$ and $L_{2}$ so it is used to define the boundary where stars are dynamically belong to the GC.
The variation of $r_{\rm h}$ is mainly driven by the internal dynamics of GCs, while $r_{\rm J}$ is mainly changed by the external dynamics from host galaxies.

\citet{Gieles+2008} find that if $\Re>\Re_{c}$ star clusters can be treated in the `tidal regime', while for $\Re<\Re_{c}$ they can be treated in the `isolated regime', where $\Re_{c}$ is a criteria to divide into the  `tidal regime' and the `isolated-regime'.
\citet{Gieles+2008} found that $\Re=0.05$, but we treat $\Re_{c}$ as a free parameter.
In our model, we assume GCs staying in the tidal regime longer than the fixed number of $t_{\rm rh}$ ($T_{\rm dur}$) are totally destroyed, where $T_{\rm dur}$ is the duration time when $\Re$ is larger than $\Re_{c}$.
That is, $T_{\rm dur}$ determines how long GCs are affected by the tidal force from their host galaxies.
We find that $T_{\rm dur}$ cannot change our results significantly so we use a fixed value of $T_{\rm dur}=0.5$~Gyrs.

The evolution of $r_{\rm h}$ is described as follows.
During the first $t_{\rm rh}$, the stellar evolution is the main driver to make GCs expand so $r_{\rm h0}$ of GCs expands $\sim$1.3 times during the first $t_{\rm rh}$ by the stellar evolution.
During the pre-core-collapse stage of GCs, we assume the 1.3~$r_{\rm h0}$ is not changed so 1.3~$r_{\rm h0}$ is maintained until the first $t_{\rm cc}$, where $t_{\rm cc}$ is the core collapse timescale of $\approx 10~t_{\rm rh}$ \citep[e.g.][]{Spitzer1987,Gurkan+2004}.
After the first $t_{\rm cc}$, because the post-core-collapse expansion by binary driven is dominant, $r_{\rm h}$ expands again \citep{Goodman1984,Spitzer1987,Baumgardt+2002,Shin+2013}:
\begin{equation}
    \label{eq9}
    r_{\rm h}(t) = r_{\rm h0}(t/t_{\rm cc})^{(2+\nu)/3},
\end{equation}
where $\nu\approx0.1$ \citep[e.g.][]{Goodman1984}.

The evolution of $r_{\rm J}$ is described as follows.
If we assume GCs are mainly affected by the tidal force of their host DM halos, $r_{\rm J}(t)$ is
\begin{equation}
    \label{eq10}
    r_{\rm J}(t) = (M/2M_{g})^{1/3}R_{g},
\end{equation}
where $M_{g}$ is the enclosed mass of host DM halos where GCs are located, and $R_{g}$ is the galactocentric radius.
Because we can trace the position of GCs in their host DM halos with time, we can calculate $R_{g}$ as a function of time.
To calculate $M_{g}$, we use the halo concentration and the virial radius in the results by the {\sc rockstar} DM halo finder, which assumes the NFW profile of DM halos \citep{Navarro+1996,Jimenez+2003,Mo+2010}.
When we calculate $r_{\rm J}$ we assume that only DM particles contribute to the $M_{g}$ so we do not consider the mass of stellar and gas components.
Contribution of stellar and gas components to $M_{g}$ will be included in a follow-up study \citep[e.g.,][]{Li+2014,Choksi+2018}.

\begin{figure}[ht]
    \plotone{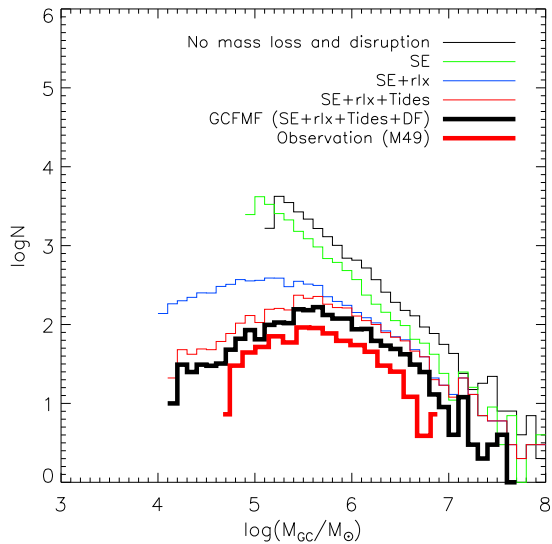}
    \caption 
    {
    The dynamical evolution of GCMF in the BCG according to recipes for the evolution and the disruption of GCs: the stellar evolution (SE), two-body relaxation (rlx), tides, and dynamical friction (DF).
    The black thin histogram shows the original GCMF without any mass-loss and disruption processes.
    The green, blue, and red thin histograms show the variation of the GCMF after applying SE, SE + tlx, and SE + rlx + tides, respectively.
    The black thick histogram represents the final GCMF after SE + rlx + tides + DF processes are applied.
    The red thick histogram shows the GCMF in M49.
    }
    \label{fig1}
\end{figure}

Finally, we can calculate $r_{\rm h}$ and $r_{\rm J}$ with time using Equations (\ref{eq9}) and (\ref{eq10}).
We assume GCs that satisfy $\Re>\Re_{c}$ longer than $T_{\rm dur}=0.5$~Gyrs are totally destroyed.
The thin red histogram in Figure \ref{fig1} shows the MF after applying tides together with the stellar evolution and two-body relaxation.
Comparing with the blue histogram, the number of low-mass GCs are decreased due to the tidal force by their host DM halos.

Some GCs are gradually moving into the center of DM halos because of dynamical friction, described as `orbital decay'.
We assume GCs that destroyed by orbital decay can contribute to the formation of nuclear star clusters (NSCs) \citep[e.g.][and references therein]{Antonini+2012,Gnedin+2014,Sanchez-Janssen+2019}.
If GCs are in the circular orbit and only the DM contributes to the velocity dispersion of GCs, the dynamical friction timescale \citep{Binney+2008,Gnedin+2014} is
\begin{equation}
    t_{\rm df} =
    0.65~{\rm Gyr}\left(\frac{R_{g}}{\rm kpc}\right)^{2}
    \left(\frac{V_{c}(R_{g})}{\rm km~s^{-1}}\right)
    \left(\frac{M(t)}{10^{5}~M_{\odot}}\right)^{-1}f_{\epsilon},
\end{equation}
where $V_{c}(R)$ is a circular velocity of the DM halo at $R_{g}$, and $f_{\epsilon}=0.5$ \citep{Gnedin+2014}.
Here, we assume $V_{c}\sim\sigma$, where $\sigma$ is the velocity dispersion of the host DM halo.
We calculate the new $t_{\rm df}$ whenever the host DM halos of GCs are changed due to DM halo mergers.
If $t_{\rm df}$ is shorter than the timescale when GCs are staying in their host DM halos, we assume GCs are destroyed and treated as NSCs.

Finally, the black thick histogram in Figure \ref{fig1} shows the final GCMF at $z=0$.
The orbital decay makes the MF slightly lower than one without it (the thin red histogram), maintaining a log-normal shape with a peak mass at $\sim$5.5$\times10^{5}$~M$_{\odot}$.
To compare our final GCMF with the observation, we over-plot the present GCMF of M49, whose $M_{\rm halo}\simeq10^{14}$~M$_{\odot}$ is similar to that of each Target 1-3 (the average $M_{\rm halo}\simeq10^{14}$~M$_{\odot}$).
The GCMF of M49 is converted from the GC luminosity function with the constant mass-to-light ratio of 2.69 in $g$-band \citep{Jordan+2007b,Willmer+2018}.
We find that the peak mass and the log-normal shape match well between our model and the observation.
Note that, if tidal shocks that we do not consider for the GC disruption are applied, the GCMF can move to the low-mass part while leaving the log-normal shape nearly invariant \citep[e.g.,][]{Fall+2001,Shin+2008,Prieto+2008,Pfeffer+2018,Li+2019,Reina-Campos+2022c}.

\begin{deluxetable*}{ccccccccc}
    \startlongtable
    \tablecaption{Parameter setting}
    \label{table1}
    \tablewidth{700pt}
    \tabletypesize{\scriptsize}
    \tablehead
    {
        \colhead{} & \colhead{$z_{\rm c}$} &
        \colhead{$\gamma_{\rm MR}$} & \colhead{initial $\log{\eta}$} &
        \colhead{$r_{\rm h0}$} & \colhead{$\Re_{c}$} &
        \colhead{$\chi^{2}$} & \colhead{$P_{\rm KS}$} &
        \colhead{$\chi^{2}$} \\
        \colhead{} & \colhead{} &
        \colhead{} & \colhead{} &
        \colhead{pc} & \colhead{} &
        \colhead{$M_{\rm GCs}$-$M_{\rm halo}$} & \colhead{GC occupancy} &
        \colhead{$f_{\rm blue}$} \\
        \cline{7-9}    
        \colhead{} & \colhead{(1)} &
        \colhead{(2)} & \colhead{(3)} &
        \colhead{(4)} & \colhead{(5)} &
        \colhead{(6)} & \colhead{(7)} &
        \colhead{(8)}
    }
    \startdata
        Fiducial & 1.0 & 0.1 & -3.3 & 3.0 & 0.05 & 0.057 & 0.96 & 0.041 \\
         & {\bf 0.0} & 0.1 & -3.3 & 3.0 & 0.05 & 0.053 & 1.00 & 0.043 \\
         & {\bf 4.0} & 0.1 & -3.3 & 3.0 & 0.05 & 0.212 & 0.00 & 0.051 \\
         & 1.0 & {\bf 0.01} & -3.3 & 3.0 & 0.05 & 0.132 & 0.81 & 0.033 \\
         & 1.0 & {\bf 0.3} & -3.3 & 3.0 & 0.05 & 0.103 & 0.63 & 0.047 \\
         & 1.0 & 0.1 & {\bf -3.0} & 3.0 & 0.05 & 0.045 & 0.94 & 0.033 \\
         & 1.0 & 0.1 & {\bf -4.0} & 3.0 & 0.05 & 0.204 & 0.04 & 0.065 \\
         & 1.0 & 0.1 & -3.3 & {\bf 2.0} & 0.05 & 0.115 & 0.10 & 0.051 \\
         & 1.0 & 0.1 & -3.3 & {\bf 10.0} & 0.05 & 0.102 & 0.02 & 0.056 \\
         & 1.0 & 0.1 & -3.3 & 3.0 & {\bf 0.5} & 0.053 & 0.98 & 0.035 \\
         & 1.0 & 0.1 & -3.3 & 3.0 & {\bf 0.005} & 0.355 & 0.00 & 0.254 \\
    \enddata
    \tablecomments
    {
        Col. (1): The lowest redshift for the GC formation.
        Col. (2): The minimum DM halo merger mass ratio.
        Col. (3): The initial mass fraction of the GC system and their host galaxies.
        Col. (4): The initial half-mass radius.
        Col. (5): The criteria of the ratio between the half-mass radius and the Jacobi radius.
        Col. (6): The chi-square value for the $M_{\rm GCs}$-$M_{\rm halo}$ relation.
        Col. (7): The K-S probability for GC occupancy.
        Col. (8): The chi-square value for the number fraction of blue GCs.
    }
\end{deluxetable*}

\section{Comparison model results with observations}
\label{sec3}

In this section, we compare our models with three representative observations at $z=0$, the $M_{\rm GCs}$-$M_{\rm halo}$ relation, the GC occupancy, and the number fraction of blue GCs ($f_{\rm blue}$) in each galaxy.
Hereafter, we use the term `galaxy' to refer to the structure, which contains $M_{\rm halo}$ in both our simulation and the observation.
To measure the similarity of our results and the observations quantitatively, we use the K-S probability ($P_{\rm KS}$) and the chi-square value:
\begin{equation}
\chi^{2}=\Sigma\frac{(x_{o}-x_{m})^{2}}{x_{o}}/N,
\end{equation}
where $x_{o}$ is the expected numbers, $x_{m}$ is the observed numbers, and $N$ is the total number of data.

To compare our results with the observation, we use two GC catalogues in this section: the HST/ACS Virgo cluster survey \citep{Cote+2004}, and the next generation Virgo cluster survey \citep[NGVS,][]{Ferrarese+2012}.
The HST/ACS Virgo cluster survey observes the 100 early-type galaxies in the Virgo cluster using the Advanced camera for Surveys on the {\em Hubble Space Telescope} in the F475W and F850LP bandpasses ($\approx$Sloan $g$ and $z$).
The wide field channel of the ACS has a field of view of 202$''\times$202$''$, which translates into 16.2~kpc$\times$16.2~kpc field of view at the distance of the Virgo cluster (16.5~Mpc).
Deep images in F475W and F850LP provide the brightest 90\% of the GC luminosity function in 100 early-type galaxies with a sample of 13,000 GCs.
The NGVS covers the Virgo cluster from its core to its virial radius (a total area of 104~deg$^{2}$) in the $u^{*}griz$ bandpasses, using the 1~deg$^{2}$ MegaCam instrument on the Canada-France-Hawaii Telescope (CFHT).
It reaches a point-source depth of $g\approx25.9$~mag (10$\sigma$) and a surface brightness limit of $\mu_{g}\sim29$~mag~arcsec$^{-2}$.

\subsection{The fiducial model}
\label{sec3.1}

\begin{figure}[ht!]
    \plotone{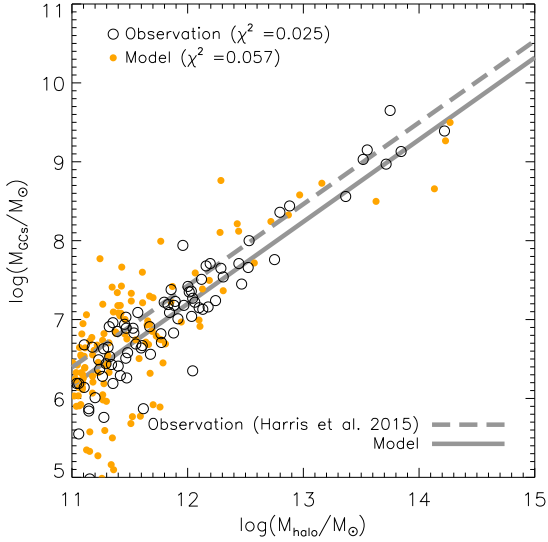}
    \caption
    {
    The $M_{\rm GCs}$-$M_{\rm halo}$ relation at $z=0$.
    Black open circles are observations of the Virgo cluster.
    Orange solid dots are results of our results with the fiducial parameter set.
    Linear fitting lines of the observations \citep{Harris+2013} and our fiducial results are represented by the dashed gray line and the dashed line, respectively.
    }
    \label{fig2}
\end{figure}

\begin{figure}[ht!]
    \plotone{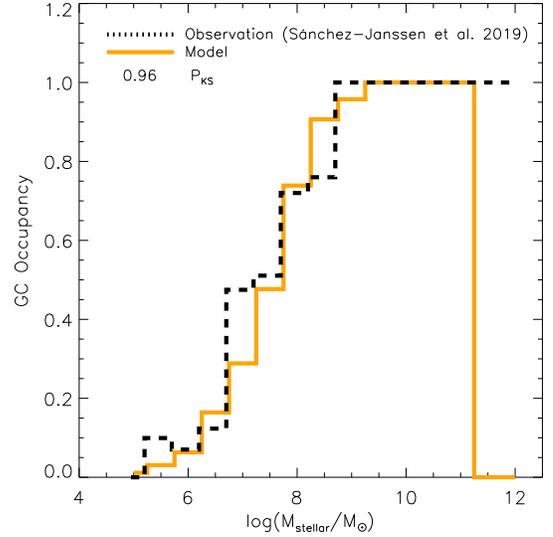}
    \caption
    {
    The GC occupancy.
    The black dashed line is the observations of the Virgo cluster \citep{Peng+2008}.
    The orange histogram shows our results with the fiducial parameter set.
    }
    \label{fig3}
\end{figure}

\begin{figure*}[ht!] 
    \plotone{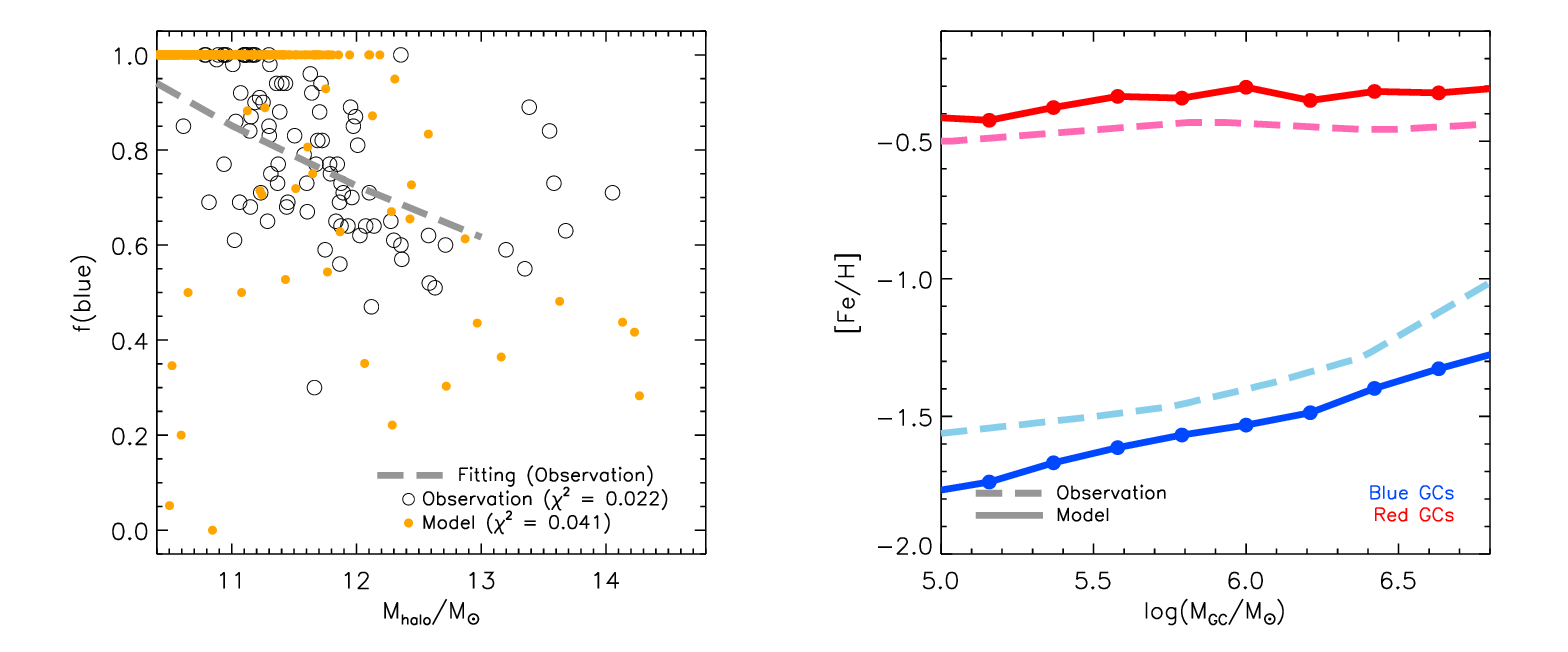}
    \caption
    {
    {\em Left}: The number fraction of blue GCs in each galaxy.
    Black open circles are observations of the Virgo cluster \citep{Peng+2008} and orange solid dots are our results with the fiducial parameter set.
    The grey dashed line shows a fitting line from the observation of the Virgo cluster \citep{Harris+2015}.
    {\em Right}: The mean metallicity of blue and red GCs (thick blue and red lines) with the GC mass.
    The coloured dashed line is a fitting line from the observation of the Virgo cluster \citep{Peng+2006,Choksi+2018}.
    We define blue GCs as [Fe/H]$<$-1.0 and red GCs as [Fe/H]$>$-1.0.
    }
    \label{fig4}
\end{figure*}

Our fiducial parameter set is determined to reproduce the overall feature of the observations, the $M_{\rm GCs}$-$M_{\rm halo}$ relation, GC occupancy, and $f_{\rm blue}$, but is not tuned to match the observations exactly.
Our fiducial values for free parameters are as below.
GCs are made until $z_{\rm c}=1$, which corresponds to the lookback time of 8~Gyr, with an estimated minimum age of GCs in the Virgo cluster \citep[e.g.,][]{Ko+2022}.
We assume GCs form from DM halo mergers of $\gamma_{\rm MR}=0.1$ with the initial $\log{\eta}=-3.3$.
We assume that $r_{\rm h0}$ is 3~pc, which is the median half-light radii of typical GCs in the Virgo cluster \citep[e.g.,][]{Jordan+2005} and GCs with $\Re_{c}>0.05$ for $T_{\rm dur}=0.5$~Gyrs are totally destroyed.
The values for fiducial parameter set is shown in the first row in Table \ref{table1}.

Figure \ref{fig2} shows the $M_{\rm GCs}$-$M_{\rm halo}$ relation of our fiducial results and the observation of the ACS Virgo cluster survey \citep{Peng+2008}.
We convert $M_{\rm stellar}$ of observed galaxies to $M_{\rm halo}$, using the $M_{\rm halo}$ to $M_{\rm stellar}$ relation \citep{Hudson+2015}.
The observation and our fiducial results have a $\chi^{2}$ of 0.025 and 0.057, respectively, but the overall trend is very similar to each other.
The mean $M_{\rm GCs}/M_{\rm halo}$ of our fiducial results is $3.5\times10^{-5}$, which is similar to the observation of $M_{\rm GCs}/M_{\rm halo} \approx 10^{-5}$ \citep{Harris+2013}.
The gray lines are fitting lines of the observation (dashed) and our fiducial results (solid).
The slopes of fitting lines of our fiducial results and the observation are 1.04 and 1.14, respectively.

Figure \ref{fig3} shows the GC occupancy (the number ratio of galaxies that contain GCs among entire galaxies) of our fiducial results compared to the observation of NGVS \citep[see their Table 4]{Sanchez-Janssen+2019}.
In both our fiducial results and the observations, most galaxies more massive than $M_{\rm stellar}$ of $10^{9}$~M$_{\odot}$ contain GCs so the overall match is good ($P_{\rm KS}\sim0.96$).
The number of low-mass galaxies that contain GCs is smaller than massive galaxies because GCs that form in low-mass galaxies are easily destroyed due to their low mass and GCs in low-mass galaxies can move to massive galaxies by galaxy mergers.

The left panel in Figure \ref{fig4} shows the number fraction of blue GCs in host galaxies ($f_{\rm blue}$).
We define blue GCs as [Fe/H]$<$-1.0 and red GCs as [Fe/H]$>$-1.0.
For comparison, we over-plot the observation of the ACS Virgo cluster survey \citep{Peng+2008}.
The grey dashed line is the fitting line of the observed data points of the Virgo cluster \citep[see][their Equation (3)]{Harris+2015}.
Values of $\chi^{2}$ of our fiducial results and the observation for $f_{\rm blue}$ are 0.041 and 0.022, respectively.
Our fiducial results show higher $\chi^{2}$ than the observation because there are many low-mass halos that contain only blue GCs \citep[e.g.][]{Peng+2008,Harris+2015}.
Red GCs can form in halos massive than $M_{\rm stellar}$ of $2.6\times10^{8}$~M$_{\odot}$ at $z=1$ by Equation (\ref{eq5}).
In our simulations, the fraction of massive halos at $z=1$ is only $\sim$0.01 so red GCs can form relatively less than blue GCs.
However, the overall behavior is similar to what is observed (increasingly red GCs as the halo mass goes high), even if the quantitative match is not very good.

The right panel in Figure \ref{fig4} shows the mean metallicity of blue and red GCs with the each GC mass ($M_{\rm GC}$) of our fiducial results and the observation of the ACS Virgo cluster survey \citep{Peng+2006,Choksi+2018}.
In observations, the mean metallicity for blue GCs increases as $M_{\rm GC}$ increases (see the blue dashed line), which is known as the `blue-tilt'.
Similar to the observation, our fiducial result (the solid line) also shows the blue-tilt \citep[e.g.,][]{Choksi+2018,Usher+2018}, but the mean metallicity of blue and red GCs tends have offsets to slightly lower and higher values of 0.1-0.3 dex.

\begin{figure*}[ht!]
    \plotone{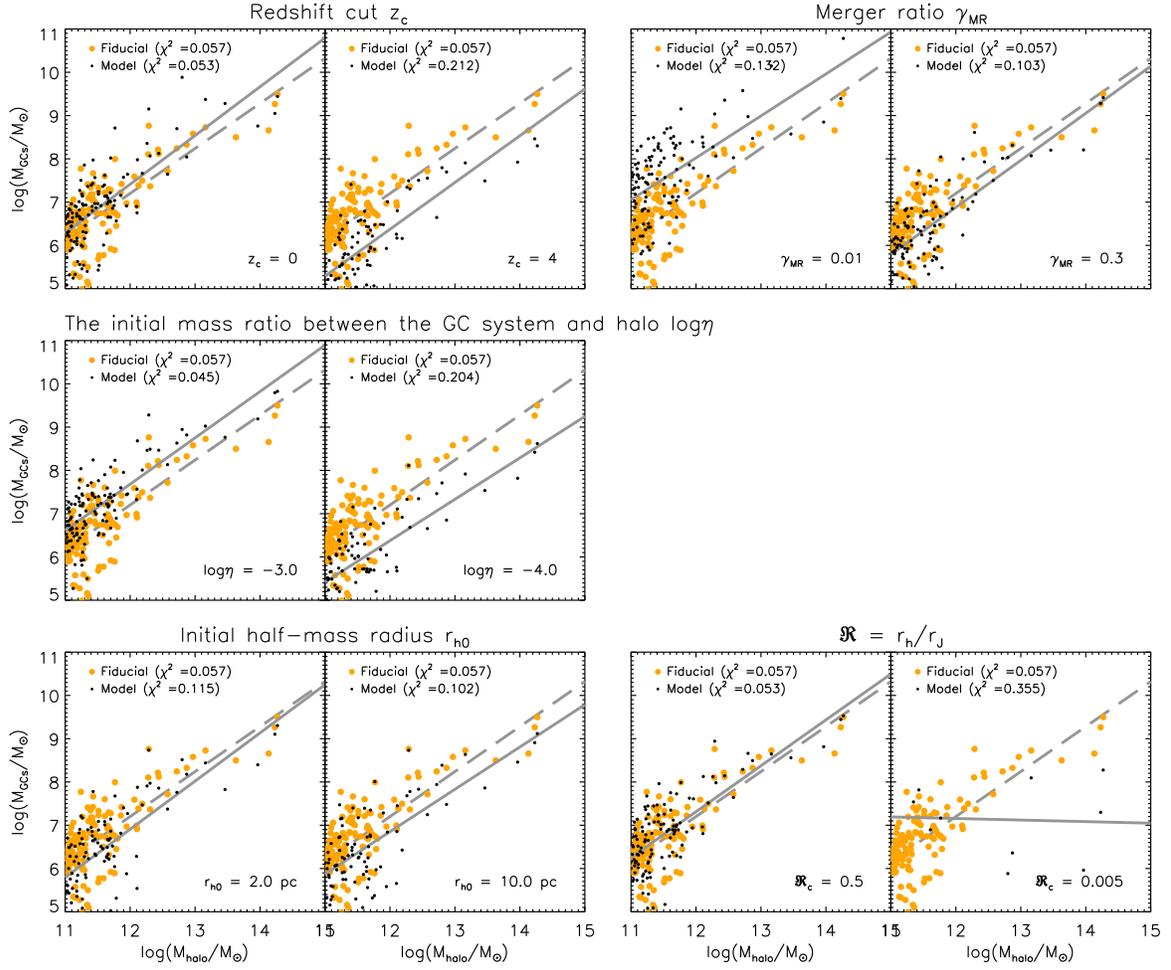}
    \caption
    {
    The $M_{\rm GCs}$-$M_{\rm halo}$ relation at $z=0$.
    Orange and black solid dots are our results with fiducial and changed parameter sets, respectively.
    Linear fitting lines of our results with fiducial and various parameter sets are represented by the dashed gray and the solid lines, respectively.
    The bottom right values are parameters that we changed from the fiducial parameter set.
    }
    \label{fig5}
\end{figure*}

\begin{figure*}[ht!]
    \plotone{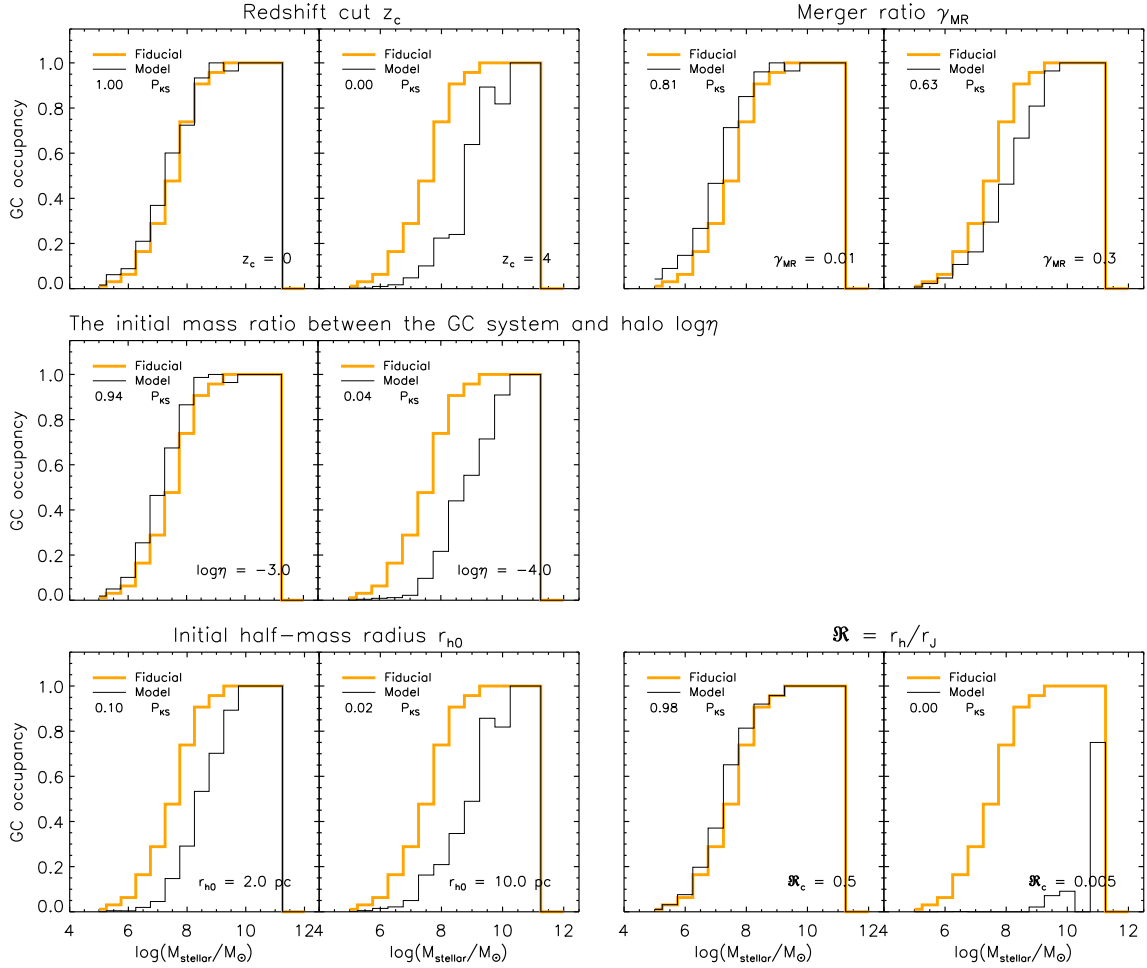}
    \caption
    {
    The GC occupancy.
    The orange and black histograms are our results with fiducial and changed parameter sets, respectively.
    The bottom right values are parameters that we changed from the fiducial parameter set.
    }
    \label{fig6}
\end{figure*}

\begin{figure*}[ht!]
    \plotone{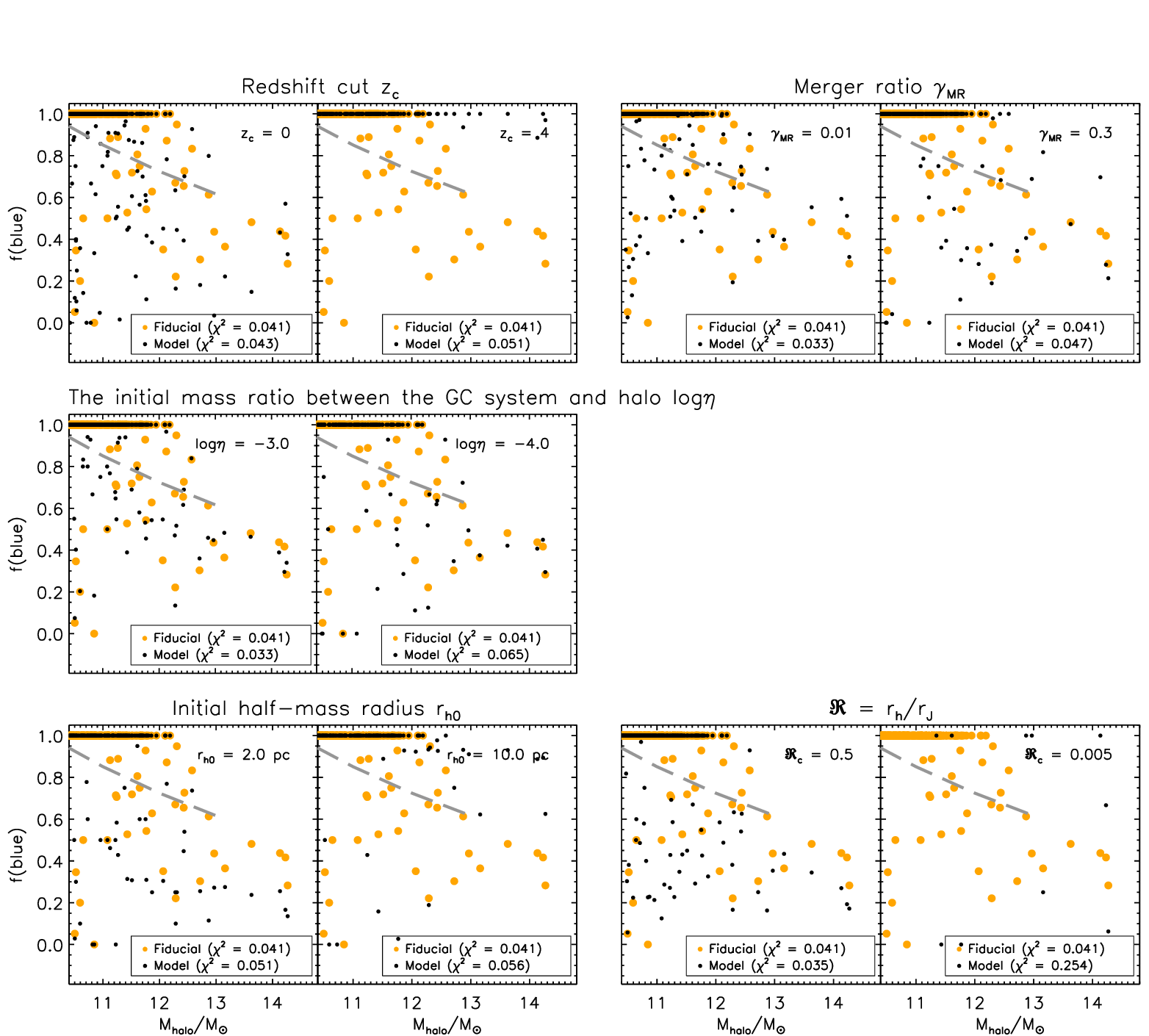}
    \caption
    {
    The number fraction of blue GCs in each galaxy.
    Orange and black solid dots are our results with fiducial and changed parameter sets, respectively.
    The gray dashed line is the fitting line from the observation \citep{Harris+2015}.
    The upper right values are parameters that we changed from the fiducial parameter set.
    }
    \label{fig7}
\end{figure*}

\subsection{Testing sensitivity to parameters}
\label{sec3.2}

In this section, we investigate how our model results are affected by the variation of free parameters.
To try to understand the importance and impact of free parameters on our results, we systematically vary each free parameter from the fiducial values (see the first row in Table \ref{table1}) while keeping the others fixed (bold numbers in Table \ref{table1} show the changed values).
Free parameters that we can change in our model are the redshift cut ($z_{\rm c}$), the merger ratio ($\gamma_{\rm MR}$), the initial mass ratio of $M_{\rm GCs}$ and $M_{\rm halo}$ ($\log{\eta}$), the initial half-mass radius of GCs ($r_{\rm h0}$), and the ratio of $r_{\rm h}$ and $r_{\rm J}$ ($\Re$).
Table \ref{table1} summarises the free parameters and the altered values we consider for them.
Note that the fiducial value is chosen to reproduce the overall observations, but is not finely tuned to match all observations exactly (see Section \ref{sec3.1}).

For comparison with our fiducial results, we use the $M_{\rm GCs}$-$M_{\rm halo}$ relation, GC occupancy, and $f_{\rm blue}$.
Figures \ref{fig5}, \ref{fig6}, and \ref{fig7} show the results of each parameter set, compared to our fiducial parameter results.

{\bf Redshift cut.}
First, we investigate how GC properties are changed with the redshift cut ($z_{c}$), which limits the minimum GC formation epoch.
The fiducial value of $z_{c}$ is 1.
However, young GCs exist in extragalactic systems \citep{Glatt+2008,Ko+2018,Usher+2019} and the mass and radii of young massive star clusters (YMCs) are similar to GCs in the local Universe \citep{Zwart+2010,Longmore+2014,Forbes+2018a}.
Thus, some simulation papers continue to make GCs until $z_{c}=0$, assuming that YMCs could be one of precursors of future GCs \citep[e.g.,][]{Prieto+2008,Li+2014,Choksi+2018,El-Badry+2019,Chen+2022}.
Applying this assumption in our model, $z_{\rm c}=1$ is changed to $z_{\rm c}=0$.
On the other hand, the average age of observed GCs in the Milky Way is old (10-13~Gyrs) so we also test using $z_{\rm c}=4$.

The results are shown in the first and second panels in the first row of Figures \ref{fig5}, \ref{fig6}, and \ref{fig7}.
In Figure \ref{fig5}, the height of $M_{\rm GCs}$ with $z_{c}=0$ is 0.30 dex higher than that of the $z_{c}=1$ case.
The slope of $z_{c}=0$ case is 1.13 similar to the that of the $z_{c}=1$ case of 1.04.
In Figure \ref{fig6}, the GC occupancy of the $z_{c}=0$ case is similar to that of the $z_{c}=1$ case ($P_{\rm KS}=1.0$).
Because both $z_{c}=0$ and $z_{c}=1$ cases experience the epoch of the maximum number of galaxy merger frequency at $z=3$, this results in the similar trend of both $z_{c}=0$ and the $z_{c}=1$ cases.
In Figure \ref{fig7}, the $z_{c}=0$ case can make more red GCs than the $z_{c}=1$ case because many red GCs form at low redshift.
Thus, the $z_{c}=0$ case has more galaxies that have lower $f_{\rm blue}$ than the $z_{c}=0$ case.

The $z_{c}=4$ case differs more strongly from the $z_{c}=1$ case because galaxies stop making GCs before the epoch of the peak galaxy merger frequency ($z=3$).
It means insufficient GCs are made overall (see Figures \ref{fig5} and \ref{fig6}) and the early formation of GCs makes galaxies have mostly blue GCs (see Figure \ref{fig7} and Equation (\ref{eq5})).
Therefore, the $z_{c}=4$ case shows a worse match than the $z_{c}=0$ case.

When we stop making GCs at $z=1$, it matches the overall observed properties of GCs in the Virgo cluster.
It implies that the GC formation in galaxy clusters after $z=1$ might be rare.
One possible explanation is that the cosmic star formation rate (SFR) has a peak at between $z=2$ and $z=1$ and the SFR is rapidly decreasing after $z=1$ in a range of $10^{11}$~M$_{\odot}<M_{\rm halo}<10^{13}$~M$_{\odot}$ \citep{Behroozi+2019}.
The other is that after galaxies fall into the galaxy cluster, they lose their gas by harassment and ram pressure stripping \citep{Moore+1996,Hester+2006,Boselli+2019}.
In this case, although the galaxy mergers continue after falling into the galaxy cluster, GCs cannot form.

{\bf Merger ratio.}
We consider the minimum galaxy merger ratio to produce the GC system ($\gamma_{\rm MR}$) is 0.1 as a fiducial value so both major mergers and minor mergers can make GCs.
We change 0.1 to 0.01 to include the extreme minor mergers and to 0.3 for major mergers only.

Results are shown in the third and fourth panels in the first row of Figures \ref{fig5}, \ref{fig6}, and \ref{fig7}.
In Figure \ref{fig5}, the $\gamma_{\rm MR}=0.01$ case makes the height of $M_{\rm GCs}$ increase 0.75 dex entirely because most galaxies undergo more frequent mergers than the $\gamma_{\rm MR}=0.1$ case.
The slopes of $\gamma_{\rm MR}=0.01$ and $\gamma_{\rm MR}=0.1$ cases are 1.04 and 0.97, respectively, so the slope stays quite similar.
More frequent galaxy mergers of the $\gamma_{\rm MR}=0.01$ case makes the GC occupancy higher than the $\gamma_{\rm MR}=0.1$ case (see Figure \ref{fig6}).
When $\gamma_{\rm MR}$ is 0.3, the height of $M_{\rm GCs}$ and the GC occupancy slightly decrease in each galaxy (see Figure \ref{fig5} and \ref{fig6}).
This is because fewer GCs form in individual galaxies due to a smaller major merger frequency than the minor merger frequency.
On the other hand, in Figure \ref{fig7}, we can see that $f_{\rm blue}$ is not significantly affected by $\gamma_{\rm MR}$ because the total number of GCs per galaxy is just determined by $\gamma_{\rm MR}$ while the number ratio of blue and red GCs is mainly decided by $z_{c}$.

{\bf The initial mass ratio between the GC system and the host galaxy.}
The initial mass ratio of the GC system and its host galaxies ($\log{\eta}$) is set to be -3.3 at GC formation epoch as our fiducial value.
We test two more initial $\log{\eta}$ of -4.0 and -3.0 \citep[e.g.][]{Choksi+2019b}.
These values assume a linear relation between $\log{M_{\rm GCs}}$ and $\log{M_{\rm halo}}$, but only the height is changed on the plot.

Panels in the second row of Figures \ref{fig5}, \ref{fig6}, and \ref{fig7} show the results.
In Figure \ref{fig5}, if the initial $\log{\eta}$ is high, more GCs can form in each galaxy so the height of $M_{\rm GCs}$ of the $\log{\eta}=-3$ case is 0.03 dex higher than or the $\log{\eta}=-3.3$ case with slopes of 1.07 and 1.04, respectively.
If the initial $\log{\eta}$ is low, insufficient GCs form so the height of $M_{\rm GCs}$ of the $\log{\eta}=-4$ case is 0.51 dex lower than the one of the $\log{\eta}=-3.3$ case.
The GC occupancy is also affected by the initial $\log{\eta}$ because the $\log{\eta}=-3$ and $\log{\eta}=-4$ cases can make more and less GCs than the $\log{\eta}=-3.3$ case (see Figure \ref{fig6}).
In Figure \ref{fig7}, the initial $\log{\eta}$ does not affect $f_{\rm blue}$ because the number ratio of blue and red GCs is mainly driven by $z_{c}$.

{\bf The initial half-mass radius.}
Next, we change the initial half-mass radius ($r_{\rm h0}$).
We use $r_{\rm h0}=3.0$~pc as a fiducial value and change 3~pc to 2~pc and 10~pc to see the effect.

The first and second panels in the third row in Figures \ref{fig5}, \ref{fig6}, and \ref{fig7} show the results.
In Figure \ref{fig5}, the slopes of the $r_{\rm h0}=2$~pc and the $r_{\rm h0}=10$~pc cases are 1.11 and 0.98, similar to the fiducial results of 1.04, but the heights of both cases are 0.22 and 0.41 dex lower than the $r_{\rm h0}=2$~pc case.
Both the $r_{\rm h0}=2$~pc and $r_{\rm h0}=10$~pc cases produce lower GC occupancy than the $r_{\rm h0}=3$~pc case entirely (see Figure \ref{fig6}).
This means too many GCs are being disrupted, but the main disruption process is different between the $r_{\rm h0}=2$~pc and the $r_{\rm h0}=10$~pc case.
If $r_{\rm h0}$ is 2~pc, two-body relaxation is the dominant process, because $t_{\rm rlx}$ is shorter than the $r_{\rm h0}=3$~pc case ( see Equation (\ref{eq7})).
But in the $r_{\rm h0}=10$~pc case, the tidal force from host galaxies is the dominant process because most GCs are in the tidal regime by the criteria of $\Re>\Re_{c}$.
Even though more GCs are being destroyed, they are not changing the overall $M_{\rm GCs}$-$M_{\rm halo}$ trend.
Instead, data points are simply being removed from the trend so the occupancy falls.

In Figure \ref{fig7}, if $r_{\rm h0}$ is 2~pc, there are more red GCs than in the $r_{\rm h0}=10$~pc case.
It is because, if $r_{\rm h0}$ is 2~pc, blue GCs that form at high redshift are destroyed by short $t_{\rm rlx}$.
In the case of red GCs, they have formed relatively recently rather than blue GCs, so red GCs can survive until $z=0$ in spite of short $t_{\rm rlx}$.
When $r_{\rm h0}$ is 10~pc, most blue and red GCs have larger $\Re$ than $\Re_{c}$ so they are quickly destroyed by tides from host galaxies.
Although GCs have $r_{\rm h0}=10$~pc, some blue and red GCs that can have low $\Re$ might be isolated from the tides.
Therefore, it make isolated GCs have a long $t_{\rm rlx}$ so they can survive until $z=0$.
These results are also seen in N-body simulations of individual star clusters \citep{Webb+2014,Zonoozi+2016,Park+2018}.

{\bf The ratio of the half-mass radius to the tidal radius.}
We use an $\Re_{c}$ of 0.05 \citep[e.g.][]{Gieles+2008} to divide the GCs into the `tidal-regime ($\Re>\Re_{c}$)' and the `isolated-regime ($\Re<\Re_{c}$)', so GCs in the tidal-regime are destroyed by the tidal force from host galaxies.
To investigate how $\Re_{c}$ affects the disruption of GCs, we change $\Re_{c}$ from 0.05 to 0.5 and 0.005.

The results are shown in the third and fourth panels in the third row in Figures \ref{fig5}, \ref{fig6}, and \ref{fig7}.
The trend of $\Re$ of each GC is increasing due to Equation (\ref{eq9}), although there is a fluctuation of $\Re$ by $r_{\rm J}$.
That is, GCs that have $\Re_{c}=0.5$ have already experienced the $\Re_{c}=0.05$ regime.
This results in similar trends of the $M_{\rm GCs}$-$M_{\rm halo}$ relation and the GC occupancy between the $\Re_{c}=0.5$ and $\Re_{c}=0.05$ cases (see Figures \ref{fig5} and \ref{fig6}).

In Figure \ref{fig7}, many galaxies have lower $f_{\rm blue}$ in the $\Re_{c}=0.5$ case than the $\Re_{c}=0.05$ case.
Because the tides are weaker in the $\Re_{c}=0.5$ case compared to the $\Re_{c}=0.05$ case, more red GCs can survive in the $\Re_{c}=0.5$ case.
However, most GCs with $\Re_{c}=0.005$ are destroyed because they are much more affected by a tidal force from their host galaxies.
As a result, none of the results with $\Re_{c}=0.005$ are similar to the $\Re_{c}=0.05$ case.
Therefore, increasing $\Re_{c}$ does not do much but decreasing this value is very significant for our model.

\section{Additional comparison with the observations}
\label{sec4}
\subsection{Comparing GC positions with observations}

\begin{figure}[ht!]
    \plotone{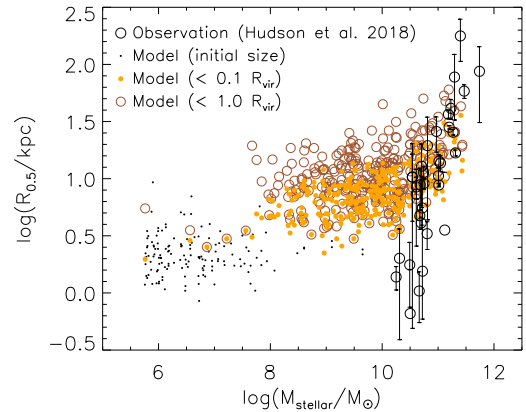}
    \caption
    {
    Median radii of the GC system in each galaxy.
    Black open circles are the observation of the Virgo cluster with errors.
    Orange solid dots and black open circles are our result with the fiducial parameter set at $z=0$, while black solid dots represent the initial $\log{R_{0.5}}$ (see the text for details).
    }
    \label{fig8}
\end{figure}

Because we use the PTM, we can trace the positions of tagged particles as GCs with time.
In this section, we focus on additional observations for comparison, using GC position information.
To compare our results with the observation, we add two more GC catalogues in this section: the point source catalogue in the SDSS Sixth Data Release \citep{Adelman-McCarthy+2008} and the Canada-France-Hawaii Telescope Legacy Survey \citep[CFHTLS,][]{Hudelot+2012}.
\citet{Lee+2010} select brightest GC candidates in a circular field with a radius of 9$^{\circ}$, using the photometry of the point sources in the SDSS Sixth Data Release.
They use the criteria for color and magnitude, $0.6<(g-i)_{0}<1.3$ and $19.5<i_{0}<21.7$~mag, with reddening correction and the magnitude limit is $i_{0}<21.7$~mag.
CHFTLS covers 155~deg$^{2}$ across four patches that comprise several fields with five filters: $u^{*}$, $g^{'}$, $r^{'}$, $i^{'}$, and $z^{'}$.
For a point source, the limiting magnitude in the $i$ band is $\sim$24.7.

\begin{figure}[ht!]
    \plotone{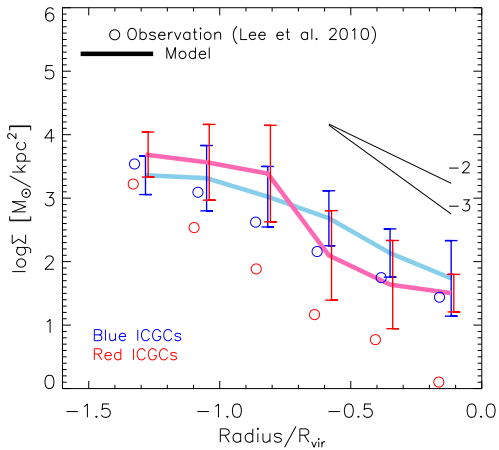}
    \caption
    {
    Mean projected density profiles of blue and red ICGCs (blue and red colors).
    Thick lines are our results with the fiducial parameter set and open circles are the observation from the Virgo cluster.
    }
    \label{fig9}
\end{figure}

\begin{figure*}[ht!]
    \plotone{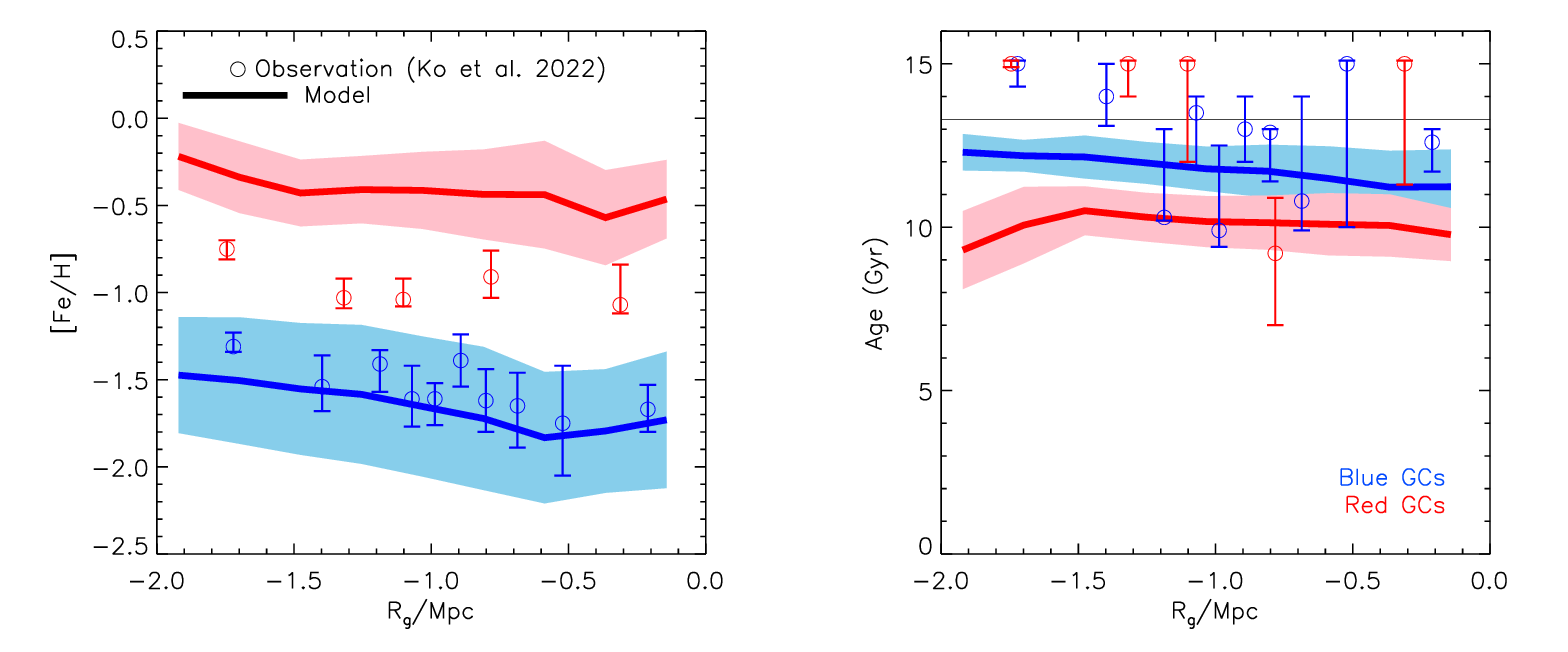}
    \caption
    {
    The mean metallicity (left panel) and the mean age (right panel) of GCs as a function of a clustercentric radius.
    Blue and red thick lines are blue and red GCs of our results with the fiducial parameter set.
    The scatter in our model distribution (1$\sigma$ errors) is shown by the shaded region.
    Open blue and red circles are blue and red GCs of the observation.
    Black line shows a maximum age of snapshots in our simulation.
    }
    \label{fig10}
\end{figure*}

First, we investigate the size of the GC system in each galaxy.
Figure \ref{fig8} shows median radii of the GC system ($\log{R_{0.5}}$) with $M_{\rm stellar}$ of their host galaxies.
The observed effective radii of the GC system are taken from nearby galaxy groups \citep{Hudson+2018}.
To measure the median radii of the GC system, we use galaxies that contain more than 10 GCs inside $0.1R_{\rm vir}$ (orange solid dots) or $1.0R_{\rm vir}$ (brown open circles).
Adopting galaxies that have more than 10 GCs can improve the statistics because using 10 GCs have higher confident to calculate $\log{R_{0.5}}$.
When we increase the radial cut from $0.1R_{\rm vir}$ to $1.0R_{\rm vir}$, $\log{R_{0.5}}$ slightly increases but the overall trend is almost the same.
The trend of $\log{R_{0.5}}$ is found to increase with $M_{\rm stellar}$ and it is similar to the observations of galaxies wih $M_{\rm stellar}>$10$^{10}$~M$_{\odot}$, although there are large vertical scatters in the observation.
From $M_{\rm stellar}=10^{8}$~M$_{\odot}$ to $M_{\rm stellar}=10^{10}$~M$_{\odot}$, the trend of $\log{R_{0.5}}$ is almost constant.
Thus, the trend of $\log{R_{0.5}}$ shows {\em a broken-power law} with a broken point at $\sim$5$\times10^{10}$~M$_{\odot}$.

We also over-plot the initial $\log{R_{0.5}}$ with black dots in Figure \ref{fig8} to show how the GC system size is changed from their initial size.
In our model, it is difficult to define the initial $\log{R_{0.5}}$ in each galaxy because GCs are added whenever galaxies experience mergers.
Thus, we just over-plot the initial $\log{R_{0.5}}$ at the first merger in each galaxy for simplicity.
We find that the initial $\log{R_{0.5}}$ is an extension of the current $\log{R_{0.5}}$.

Next, we investigate the distribution of ICGCs in galaxy clusters.
Various galaxy cluster surveys have detected ICGCs and they found that the number density profile of blue ICGCs is more extended than that of red ICGCs \citep{Lee+2010,Peng+2011,Durrell+2014,Madrid+2018,Harris+2020}.
Figure \ref{fig9} shows the projected density profiles of blue and red ICGCs of our fiducial results and the Observation of the Virgo cluster \citep{Lee+2010}.
In the observation, ICGCs are defined by the masking method \citep[e.g.,][]{Lee+2010,Ko+2017,Ko+2018,Harris+2020}\footnote{To remove GCs that are members of satellite galaxies, \citet{Lee+2010} mask out a circular region with 5$R_{25}$, where $R_{25}$ is a radius of a galaxy where the surface brightness $\mu_{B}$ = 25 mag arc sec$^{-2}$.}.
In our simulation, we define ICGCs that are actual members of the galaxy cluster, rather than members of satellite galaxies using the binding energy calculated by the {\sc rockstar} DM halo finder.

Our fiducial results show that the blue ICGCs generally have a more extended surface density profile as the observation, although there is an increasing trend of red ICGCs at the outskirt of the galaxy cluster due to red ICGCs that are coming from the recent mergers of Milky Way-size galaxies.
Because blue GCs are older than red GCs, there is a high probability that blue GCs can escape from their host galaxies by accretion or merger.
It makes blue ICGCs have a higher mass density profile than red ICGCs.

In Figure \ref{fig10}, we investigate the mean metallicity (left panel) and the mean age (right panel) of GCs as a function of a clustercentric radius.
For comparison, we over-plot the NGVS observation as open circles \citep{Ko+2022}.
In the left panel of Figure \ref{fig10}, both our fiducial results and the observation show a decreasing mean metallicity of blue and red GCs with the clustercentric radius.
Especially, the gradient and the height of the mean metallicity of blue GCs match the observation well.
However, in the case of red GCs, there is a offset of the mean metallicity between our fiducial results and the observation.
We revisit this issue in the discussion (Section \ref{sec5}).
In the right penal of Figure \ref{fig10}, the mean age of blue GCs is decreasing with the clustercentric radius in both our fiducial results and the observation but there is no overlap.
However, there is a possibility that the age of GCs is estimated younger (1-2~Gyr) because integrated stellar populations from the observations can be affected by changing the horizontal branch morphology \citep{Ko+2022}.
In this case, there might be an overlap of blue GCs between our fiducial results and the observation but the age of red GCs in our fiducial model is still younger than the observation.

\begin{figure*}[ht!]
    \plotone{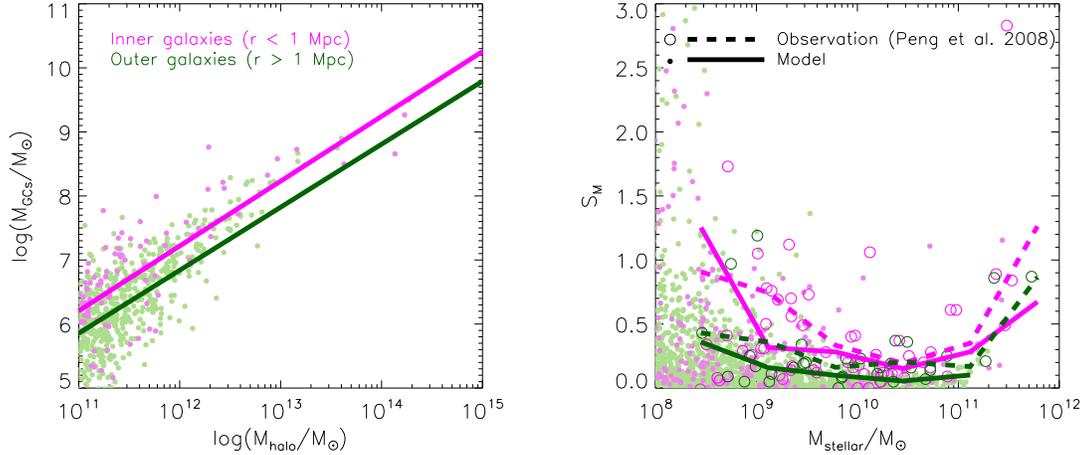}
    \caption
    {
    {\em Left: }
    The $M_{\rm GCs}$-$M_{\rm halo}$ relation of our results with the fiducial parameter set.
    Green and magenta colors represent inner and outer galaxies, respectively.
    The thick lines are linear fitted lines.
    {\em Right: }
    The specific mass of GCs with $M_{\rm stellar}$.
    The thick solid and dashed lines are our results with the fiducial parameter set and the observation of the Virgo cluster, respectively.
    Solid dots are our fiducial results and open circles are the observation.
    }
    \label{fig11}
\end{figure*}

\subsection{Properties of the GC system with the clustercentric radius}

To understand how properties of the GC system are changed by the clustercentric radius, we investigate the $M_{\rm GCs}$-$M_{\rm halo}$ relation and the specific mass \citep[$S_{\rm M}=100M_{\rm GCs}/M_{\rm stellar}$,][]{Peng+2008}. 
We divide galaxie into inner ($r<$1~Mpc) and outer galaxies ($r>$1~Mpc).

The left panel of Figure \ref{fig11} shows the $M_{\rm GCs}$-$M_{\rm halo}$ relation of our fiducial results.
The slopes of inner and outer galaxies are 1.06 and 0.94, respectively, and the height of the inner galaxies is 1 dex higher than the outer galaxies.
The right panel of Figure \ref{fig11} shows $S_{\rm M}$ of our fiducial results and the observation of the ACS Virgo cluster survey \citep{Peng+2008}.
The inner galaxies has higher $S_{\rm M}$ than the outer galaxies in both our fiducial results and the observation \citep[e.g.][]{Peng+2008,Harris+2013}.
There are insufficient high-mass and low-mass galaxies in both our simulation and the observation, respectively, so it makes a difference between our fiducial model and the observation at high-mass and low-mass regions of galaxies significantly \citep[e.g.][]{Mistani+2016,Carlsten+2021}.
However, we still can see a U-shape of $S_{\rm M}$ of the inner galaxies in both our fiducial results and the observation.

Overall, Figure \ref{fig11} shows that the inner galaxies have more GCs than the outer galaxies.
This results from that the inner galaxies generally experience more frequent mergers than the outer galaxies due to an early infall to galaxy clusters.

\section{Discussion}
\label{sec5}

\begin{figure*}
    \gridline{
             \fig{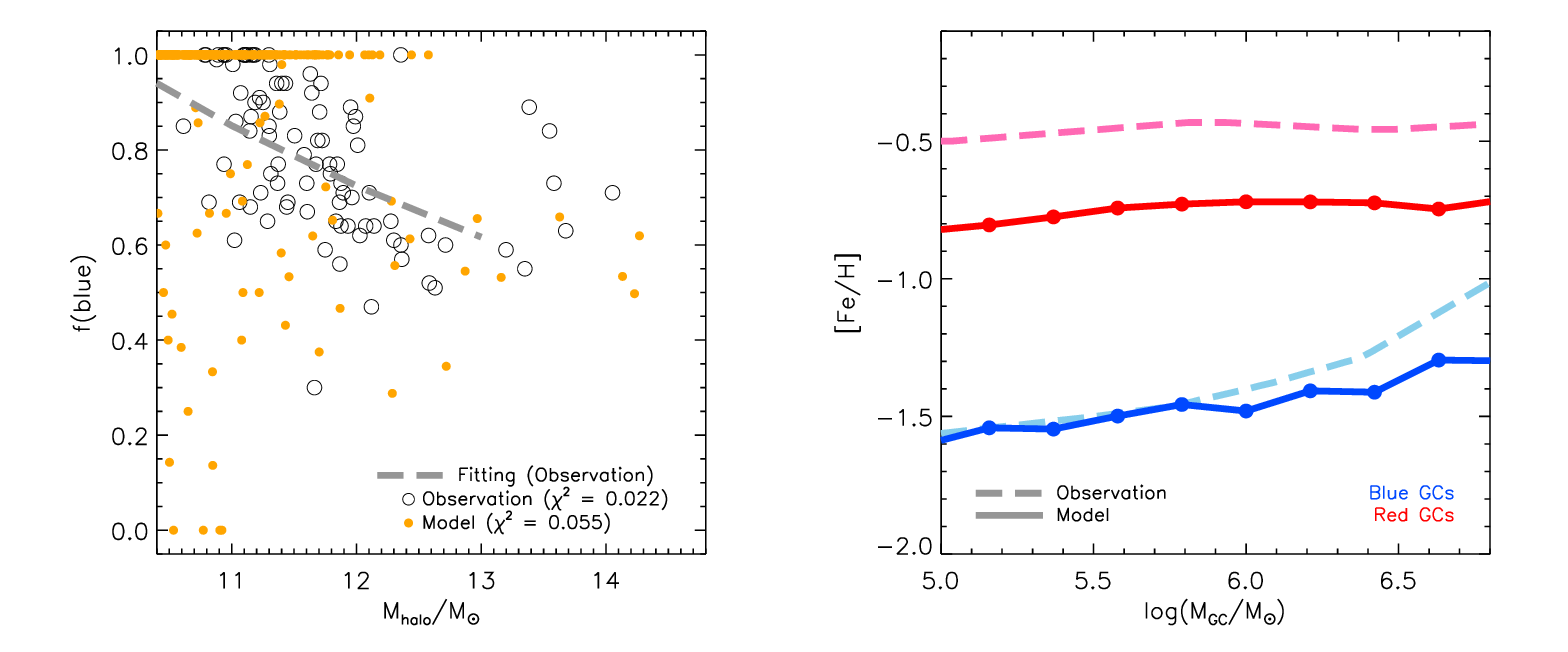}{0.8\textwidth}{ }
             }
   \gridline{
             \fig{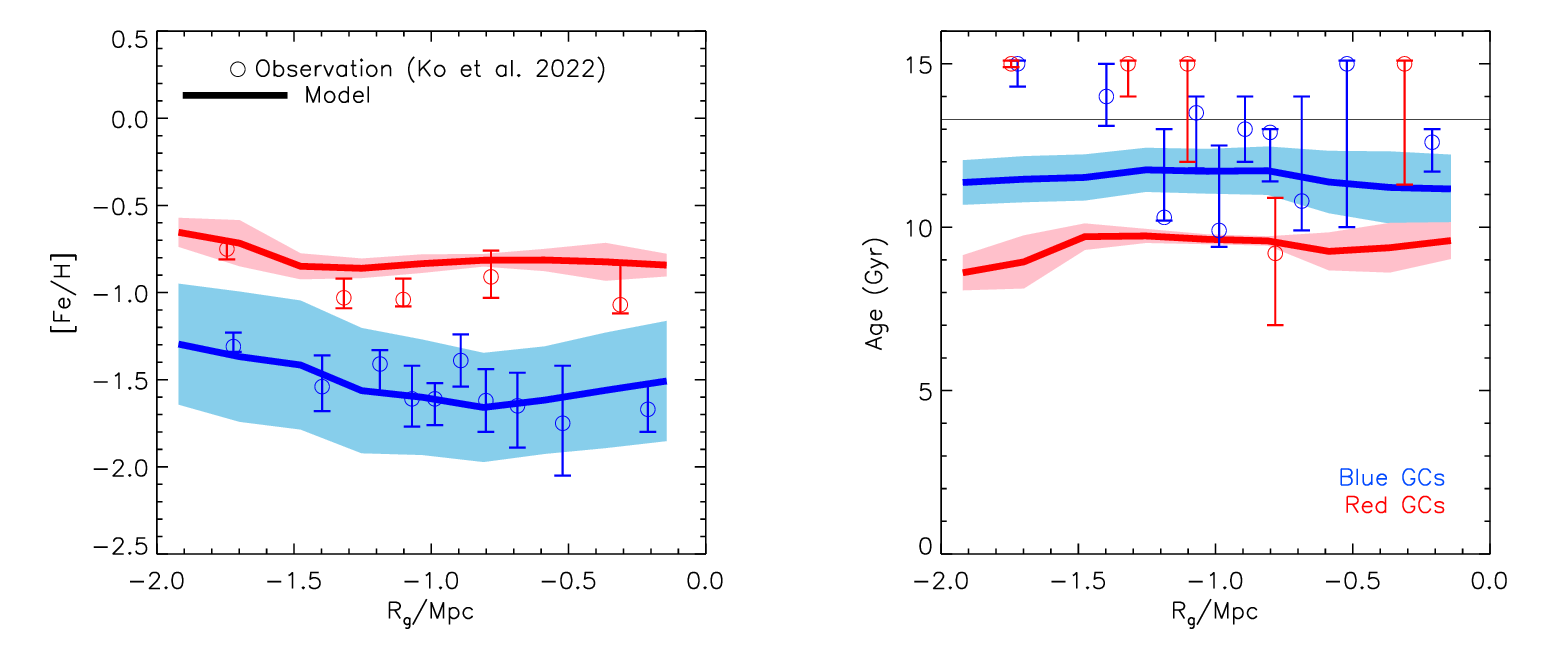}{0.8\textwidth}{ }
             }
    \caption{
    With different metallicity parameters of $\alpha_{m}=0.1$ and $\alpha_{z}=2.0$ in Equation (\ref{eq5}).
    {\em Upper left panel:} The number fraction of blue and red GCs per galaxy.
    {\em Upper right panel:} The mean metallicity of blue and red GCs with a GC mass.
    {\em Below left panel:} The mean metallicity of blue and red GCs with a clustercentric radius.
    {\em Below right panel:} The mean age of blue and red GCs with a clustercentric radius.
    Blue and red shade regions in both below left and right panels are the scatter in our model distribution (1$\sigma$ errors).
    }
    \label{fig12}
\end{figure*}

In this paper, the most important issue we try to tackle is how the various free parameters impact on the GC properties at $z=0$.
In Section \ref{sec3.2}, we investigate how the GC properties can be changed by the various values of free parameters: the redshift cut ($z_{c}$), the merger ratio ($\gamma_{\rm MR}$), the initial mass ratio between the GC system and the host galaxy ($\log{\eta}$), the initial half-mass radius ($r_{\rm h0}$), and the ratio of $r_{\rm h}$ to $r_{\rm J}$ ($\Re$).
These parameters change the $M_{\rm GCs}$-$M_{\rm halo}$ relation, GC occupancy, and $f_{\rm blue}$ significantly.
Among them, $z_{\rm c}$, $\gamma_{\rm MR}$, initial $\log{\eta}$, and $\Re$ are sensitive free parameters because they affect the $M_{\rm GCs}$-$M_{\rm halo}$ relation, GC occupancy, and $f_{\rm blue}$ simultaneously.
This means the environment of the GC formation and the tide are important to build up the current properties of GCs in galaxy clusters.
However, because we examine the effect of parameter variations by changing only one parameter value and leaving the rest of the values fixed, we do not know whether there are inter-dependencies or degeneracy among the parameters.
In addition, one parameter can affect some of the observations significantly but cannot affect the rest the observations.
For example, the $M_{\rm GCs}$-$M_{\rm halo}$ relation is less affected by $r_{\rm h0}$ but GC occupancy and $f_{\rm blue}$ are significantly altered by $r_{\rm h0}$.
Various parameter combinations are needed to derive the formation conditions for galaxies and GCs.
In the future, we propose to use the Markov chain Monte Carlo (MCMC) method to improve our understanding of how various parameter combinations can affect the final results and the match to observations.

Throughout the paper, we assume that the initial $\log{\eta}$ is a constant.
However, stars form from cold gas so previous semi-analytical models have assumed a constant fraction between $M_{\rm GCs}$ and the cold gas mass ($M_{\rm gas}$), which is a function of $M_{\rm stellar}$ and $z$ \citep[e.g.][]{Li+2014,Choksi+2018,Choksi+2019a,Choksi+2019b}.
Instead of using a constant initial fraction between $M_{\rm GCs}$ and $M_{\rm halo}$, we also tried to adopt a constant fraction between $M_{\rm GCs}$ and $M_{\rm gas}$, using the best model parameters in \citet{Choksi+2018}.
The $M_{\rm GCs}$-$M_{\rm halo}$ relation at $z=0$ of our fiducial results matches the observation, while the $M_{\rm GCs}$-$M_{\rm halo}$ relation at $z=0$ with a constant fraction between $M_{\rm GCs}$ and $M_{\rm gas}$ is an order lower than the observation of the Virgo cluster.
We infer that the different method for GC disruption by tide and the additional GC disruption, dynamical friction, which \citet{Choksi+2018} did not consider, might demand a higher initial $\log{\eta}$.
We will investigate how additional GC mass loss and disruption processes can affect the initial mass fraction of the GC system and its host galaxy.

Our fiducial results can reproduce overall GC properties in the galaxy cluster, the $M_{\rm GCs}$-$M_{\rm halo}$ relation, GC occupancy, and the decreasing $f_{\rm blue}$ with $M_{\rm halo}$.
However, our fiducial results have some limitations: the abundance of low-mass galaxies that have only blue GCs (the left panel of Figure \ref{fig4}), and the offsets of the mean metallicity of blue and red GCs between our model and the observation (the right panel of Figures \ref{fig4} and \ref{fig10}).
Despite the variations in the free parameters that we make, there is generally a much higher number of low-mass galaxies that have only blue GCs than the observation (Figure \ref{fig7}).

To improve the match, we change $\alpha_{m}=0.35$ to $\alpha_{m}=0.1$ and $\alpha_{z}=0.9$ to $\alpha_{z}=2.0$ in the metallicity model (Equation (\ref{eq5})), while keeping the free parameters as the fiducial values (see Table \ref{table1}).
Results are shown in Figure \ref{fig12}.
It is interesting that despite changing metallicity parameters to match the height of our model with the observation (see the bottom left panel), $f_{\rm blue}$ is not altered significantly (see the top left panel).
We still can see a similar trend of blue GCs between the changed model and the observation.
In the upper right panel, the mean metallicity of red GCs in the observation is 0.3 dex higher than our model.
In the case of blue GCs, the trend of the changed model is same with the observation up to $M_{\rm GC}\sim10^{6}$~M$_{\odot}$. 
In the case of massive blue GCs ($M_{\rm GC}>10^{6}$~M$_{\odot}$), there is a offset between the changed model and the observation but we still can see a blue-tilt.
In the lower right panel, we do not see a clear age gradient in the blue GCs and there is no overlap between our model and the observation.

Although we change $\alpha_{m}$ and $\alpha_{z}$ substantially in the metallicity model to reproduce the observation in the below left panel in Figure \ref{fig12}, our model still has a limitation to reproduce the GC metallcity in detail: the abundance of low-mass galaxies that have only blue GCs, the height of the mean metallicity of red GCs, and the age trend of blue and red GCs with a clustercentric radius.
We expect that this issue could be improved by changing GC formation scenarios.

In addition, we adopt the metallicity model that the metallicity is a function of $M_{\rm stellar}$ and $z$, which means the metallicity bimodality represents the age bimodality.
However, some galaxies that have metallicity bimodality do not reflect age difference \citep{Beasley+2000,Hempel+2007}.
Instead, the age-metallicity distribution of GCs at $z=0$ is useful tool to infer galaxy assembly \citep{Forbes+2010}.
We revisit assembly histories of the Virgo cluster using the age-metallicity space in the next paper.

Our model naturally produces the GC color bimodality due to the hierarchical merging scenario and the metallicity model (Equation (\ref{eq5})) so our model is not proper to investigate some galaxies that do not have the GC color bimodality \citep[e.g., M31 and many ellipticals:][]{Larsen+2001}.
However, our fiducial results can reproduce overall observed GC properties in the galaxy cluster, not only the $M_{\rm GCs}$-$M_{\rm halo}$ relation, the GC occupancy, and the decreasing $f_{\rm blue}$ with $M_{\rm halo}$, but also a discrete mean metallicity between blue and red GCs.
Therefore, we suggest that our model is still useful to investigate the GC properties like the GC metallicity bimodality and the fraction of blue and red GCs in each galaxy in galaxy clusters.

In this paper, we have so far not investigated the velocities and dynamics of our modelled GCs.
But in the future, this could be a valuable application of our method as GC velocities can provides additional information for comparison with observations such as using them as a tracer to measure the DM mass of galaxies \citep[e.g.,][]{Smith+2013,Doppel+2021,Hughes+2021} and the fact that blue GCs are observed to have a higher rotation velocity and velocity dispersion than red GCs \citep{Lee+2010,Schuberth+2010,Strader+2011,Durrell+2014,Ko+2020,Chaturvedi+2021}.

In Section \ref{sec3}, we compare our results with three representative observations.
The GC occupancy is an important observation because sometimes the GC occupancy is wrong even if the $M_{\rm GCS}$-$M_{\rm halo}$ relation looks fine.
However, the sensitivity of $f_{\rm blue}$ does not seem to be higher than the other observational results because it is not changed significantly by the variation of free parameters.
As we mentioned, if we solve the limitations of the GC metallicity using other GC formation scenarios, $f_{\rm blue}$ might be an important observation, which can constrain our model to match the observation.
In addition, if we use the velocity information, it might be an important observation to constrain the best parameter set, which provides the formation environment of GCs exactly.

Due to the huge recent developments in GC observations, the positions, velocities, and other properties of (IC)GCs are now estimated in multiple other galaxy clusters besides the Virgo cluster, e.g., Coma, Abell, Perseus, Fornax clusters in cluster surveys: the HST/ACS Virgo cluster survey \citep{Cote+2004}, the NGVS \citep{Ferrarese+2012}, the HST/ACS Coma cluster survey \citep{Carter+2008}, the HST/ASC Fornax cluster survey \citep{Jordan+2007b}, and the next generation Fornax cluster survey \citep{Eigenthaler+2018}.
Using our PTM with the semi-analytical approach, we are in an excellent position to compare our models with various state-of-the-art observations at low redshifts.
As a result, we can study variations in environmental properties or conditions that can help to reproduce the current properties of GCs and their host galaxies.

\section{Summary}
\label{sec6}

We investigate the properties between GCs and host galaxies in galaxy clusters, using the cosmological zoom-in simulations for the Virgo cluster.
We use a particle tagging method with the semi-analytical approach, assuming the hierarchical merging scenario: GCs form from galaxy mergers and their metallicity is assigned based on the stellar mass of host galaxies and formation redshift of GCs.
We apply the internal and external mechanisms to the evolution of each GC: stellar evolution, two-body relaxation, tides, and dynamical friction.
Using the semi-analytical approach, the formation and evolution of GCs are controlled by free parameters.
The main goal of the paper is not to reproduce the observations quite well but to test the sensitivity to physical processes for the GC formation.
Our results are summarized below.
\begin{enumerate}
    \item Our fiducial parameter set can reproduce not only the $M_{\rm GCs}$-$M_{\rm halo}$ relation but also GC occupancy, the decreasing $f_{\rm blue}$ with $M_{\rm halo}$, and the blue-tilt.
    However, our fiducial parameter set has a limitation to reproduce the observed GC metallicity in detail: the abundance of low-mass galaxies that have only blue GCs (the left panel of Figure \ref{fig4}), the mean metallicity (the right panel of Figure \ref{fig4} and the left panel of Figure \ref{fig10}), and the age trend of the blue and red GCs (the right panel of Figure \ref{fig10}).
    
    \item Among free parameters, $z_{\rm c}$, $\gamma_{\rm MR}$, initial $\log{\eta}$, and $\Re$ are important parameters because they affect the $M_{\rm GCs}$-$M_{\rm halo}$ relation, GC occupancy, and $f_{\rm blue}$, simultaneously.
    These parameters affect the formation and evolution of GCs so we will investigate environment of GC formation and evolution using the MCMC method.
    
    \item The position information, traced by the PTM, makes it possible for us to compare with additional observations: the GC system size in each galaxy ($\log{r_{0.5}}$), the projected density profiles of blue and red ICGCs, the metallicity and age gradient with the clustercentric radius, and the radial dependence of the specific frequency ($S_{\rm M}$).
    However, there are offsets of the mean metallicity and age of red GCs between our fiducial model and the observation of the Virgo cluster (Figure \ref{fig10}).
\end{enumerate}
In recent times, large samples of (IC)GCs in various galaxy clusters have been observed, therefore the demand for modelling that attempts to simultaneously reproduce multiple aspects of their properties has never been higher.
In the future, we plan to use our model to investigate the GCs in various galaxy clusters, to understand how their properties depend on their formation environment such as cluster mass, cluster dynamical state, and cluster merger history. 


\begin{acknowledgments}
We deeply thank the anonymous referee for the helpful comments that improved the quality of the paper.
This research was supported by the Korea Astronomy and Space Science Institute under the R\&D program (Project No. 2022-1-830-06) supervised by the Ministry of Science and ICT.
J.S. acknowledges support from the National Research Foundation of Korea grant (2021R1C1C1003785) funded by the Ministry of Science, ICT and Future Planning.
K.C. was supported by the National Research Foundation of Korea (NRF) grant funded by the Korea government (MSIT) (2021R1F1A1045622).

\end{acknowledgments}

%




\appendix
\section{The Schechter initial GCMF}
\begin{figure*}[ht!]
    \includegraphics[width=\textwidth]{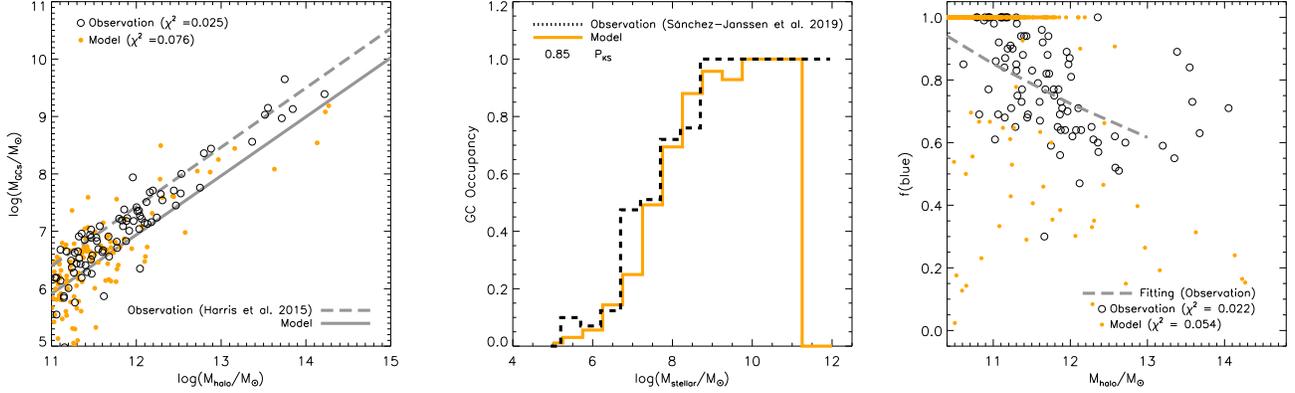}
    \caption
    {
    Our fiducial model with a Schechter initial GCMF.
    The truncation mass is $10^{6}$~M$_{\odot}$ and the mass range of GCs is $10^{5}$-$10^{8}$~M${_\odot}$.
    }
    \label{figa1}
\end{figure*}

Our model adopts the power-law initial GCMF.
To see the effect of the initial GCMF, we test our model with the Schechter initial GCMF \citep{Schechter1976,Gieles2009,Adamo+2020}.
Figure \ref{figa1} shows the $M_{\rm GC}$-$M_{\rm halo}$ relation, GC occupancy, and $f_{\rm blue}$.
We find that there is no large difference between the power-law and Schechter initial GCMFs.

\section{The effect of changing the minimum mass of GCs}
\begin{figure*}[ht!]
    \includegraphics[width=\textwidth]{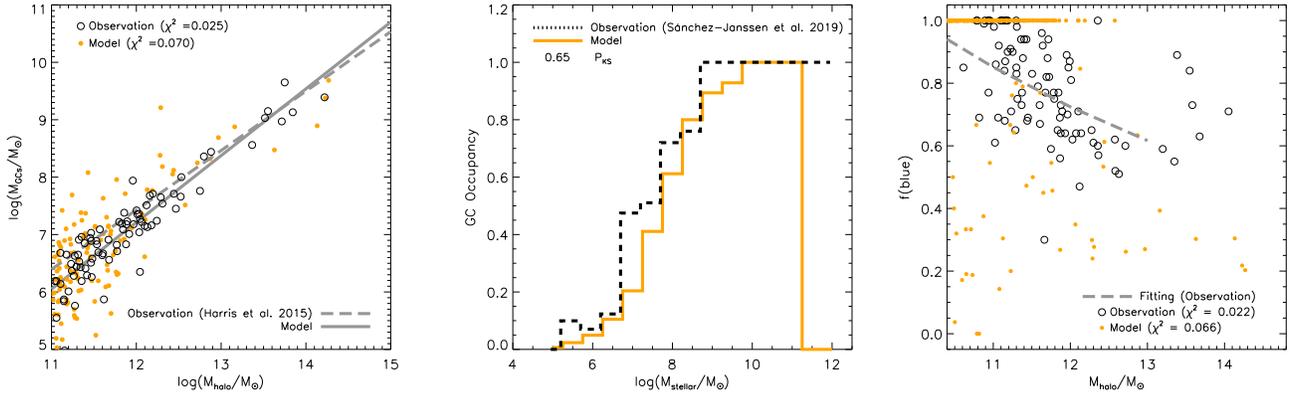}
    \caption
    {
    Our fiducial model with $M_{\rm min}=10^{4}$.
    }
    \label{figa2}
\end{figure*}

Our model adopts the minimum mass of GCs ($M_{\rm min}$) as $10^{5}$~M$_{\odot}$ because we assume that the low-mass GCs are destroyed quickly by two-body relaxation.
To investigate the effect of $M_{\rm min}$, we change $10^{5}$~M$_{\odot}$ to $10^{4}$~M$_{\odot}$.
Figure \ref{figa2} shows the $M_{\rm GC}$-$M_{\rm halo}$ relation, GC occupancy, and $f_{\rm blue}$.
We fine that results are not significantly different with $M_{\rm min}=10^{5}$~M$_{\odot}$.
In addition, we use GCs whose mass is higher than $10^{4}$~M$_{\odot}$ at z=0 to analyze our results because the observed mass range of GCs is $10^{4}$-$10^{6}$~M$_{\odot}$.
Thus, we suggest that $M_{\rm min}$ does not affect the results significantly.

\bibliography{ICGC_01}{}
\bibliographystyle{aasjournal}



\end{CJK*}
\end{document}